\documentclass[final,3p]{elsarticle}
\usepackage{graphicx}
\usepackage{amsmath}
\usepackage{amsfonts}
\usepackage{mathrsfs}
\usepackage[OT1]{fontenc} 
\usepackage{enumerate}
\usepackage{amsthm}
\usepackage{tabularx}
\usepackage{arydshln}

\usepackage[colorlinks=true,linkcolor=blue,citecolor=blue]{hyperref}
\usepackage{mwe}
\usepackage{subfig}
\usepackage{float}
\usepackage{graphicx}
\usepackage{textcomp}
\usepackage{xcolor}
\usepackage{footnote}
\usepackage{makecell}
\usepackage{graphicx}
\usepackage{amsmath}
\usepackage{amsfonts}
\usepackage{mathrsfs}
\usepackage[OT1]{fontenc} 
\usepackage{enumerate}
\usepackage{amsthm}
\usepackage{tabularx}
\usepackage{arydshln}
\usepackage{fixfoot}
\colorlet{re}{black}
\colorlet{r2}{black} 
\colorlet{r3}{red} 
\usepackage{colortbl}
\usepackage{tikz}

\usepackage{subfig}
\usepackage{textcomp}
\usepackage{xcolor}
\usepackage{footnote}
\usepackage{makecell}
\usepackage[draft]{todonotes}
\usepackage[section]{placeins}

\usepackage{algorithm}
\usepackage{algpseudocode}
\usepackage[inline]{enumitem}
\usepackage{mathtools}
\usepackage{lineno}
\colorlet{r1}{black}
\colorlet{r2}{black}

\biboptions{sort&compress}

\usepackage{pifont}
\color[rgb]{0,0,0}

\colorlet{add}{black}

\begin{document} 

\begin{frontmatter}

\title{Traffic Flow Characteristics and Lane Use Strategies for Connected and Automated Vehicle in Mixed Traffic Conditions}


\author[1]{Zijia Zhong\corref{cor1}}
\address[1]{John A. Reif, Jr. Department of Civil and Environmental Engineering, New Jersey Institute of Technology,\\ United States}
\cortext[cor1]{Corresponding author: Zijia Zhong, zijia.zhong@njit.edu \newline
\textbf{Please cite this article: Z. Zhong, J. Lee, and L. Zhao ``Traffic flow characteristics and lane use strategies for connected and automated vehicles in mixed traffic conditions'', J. Advanced Transp., vol 2021, doi: 10.1155/2021/8816540}}

\author[1]{Joyoung Lee}

\author[2]{Liuhui Zhao}
\address[2]{College of Engineering, Georgia Institute of Technology, Atlanta, GA 30332}

\begin{abstract}
\textcolor{r2}{Managed lanes, such as a dedicated lane} for connected and automated vehicles (CAVs), can provide not only technological accommodation but also desired market incentives for road users to adopt CAVs in the near future. \textcolor{r1}{In this paper, we investigate traffic flow characteristics with two configurations of the managed lane across different market penetration rates and quantify the benefits from the perspectives of lane-level headway distribution, fuel consumption, communication density, and overall network performance.} The results highlight the benefits of implementing managed lane strategies for CAVs: 1) a dedicated CAV lane significantly extends the stable region of the speed-flow diagram and yields a greater road capacity. As the result shows, the highest flow rate is 3,400 vehicles per hour per lane at 90\% market penetration rate with one CAV lane; 2) the concentration of CAVs in one lane results in a narrower headway distribution (with smaller standard deviation) even with partial market penetration; 3) a dedicated CAV lane is also able to eliminate duel-bell-shape distribution that is caused by the heterogeneous traffic flow; and 4) a dedicated CAV lane creates a more consistent CAV density, which facilitates communication activity and decreases the probability of packet dropping.
\end{abstract}
  

\end{frontmatter}


\section{Introduction}
\textcolor{r2}{
The mobility landscape is experiencing a paradigm shift due to rapid advancements of the information and vehicular technologies. Among them, the connected and automated vehicle (CAV) technologies have been contributing to the adoption of next-generation vehicles that are equipped with connectivity (i.e., connected vehicles) and/or automation (i.e. automated vehicles).
}
In spite of CAV's immense benefits and potentials in reshaping the mobility landscape, the adoption of CAVs by consumers is still uncertain \cite{chan2017advancements}, although some lower-level vehicle automation in the form of driver-assistance system has been commercially available. 

The near-term deployment of CAVs is characterized by mixed traffic conditions, where human-driven vehicles (HVs) and CAVs constantly interact with each other. 
\textcolor{r2}{
The potential benefits from CAVs may be offset by the interactions among different types of vehicles. For example, the short following time gap (e.g., 0.6 $s$) is only feasible when a CAV follows another CAV. To overcome such shortcoming in near-term CAV deployment, managed lane strategies, such as CAV dedicated lane, is one of the promising solutions in order to facilitate the formation of the CAV strings. Practically, managed lane strategies are freeway lanes that are set aside and operated under various fixed and/or real-time strategies in response to certain objectives, such as improving traffic operation \cite{fhwa2008Managed}. It is also anticipated that managed lane strategies incentivize the adoption of CAV, just as they did for encouraging car-pooling or low emission vehicles.}
%

\textcolor{r1}{The goal of this study is to investigate the impact of different lane use strategies under mixed traffic conditions at vehicle trajectory- as well as lane-level. For clarity, we refer mixed traffic condition to the condition that CAVs and HVs operate on the same roadway network in the following discussions. The contributions of the paper include:}
\begin{enumerate}
    \item the analysis of CAV-enhanced traffic flow characteristics at the lane- and vehicle-level,
    \item the investigation of traffic performance with gradual introduction of CAV platoons under difference managed lane strategies, and 
    \item the implications of managed lane strategies from a dedicated short-range communication (DSRC) communication perspective.
\end{enumerate}

The remainder of the paper is organized as follows. Related work regarding the research of CAVs in mixed traffic and managed lanes is reviewed in Section \ref{sect:Literature}, followed by the evaluation methodology, including customized CAV module and defined scenarios, in Section \ref{sect:methodology}. The simulation results are presented and discussed in Section \ref{sect:result}. Lastly, findings and recommendations are discussed in Section \ref{sect:conclusion}.



\section{Literature Review}
\label{sect:Literature}

\textcolor{r1}{
There have been numerous studies on the implementation and evaluation of CAVs in various traffic settings. Aligning with our research topic, we focused our literature search on two key aspects of CAV studies: 
\begin{enumerate*}[label=\arabic*)]
\item CAV evaluation in mixed traffic conditions at network level and
\item managed lane strategies for CAV.
\end{enumerate*}
}

\subsection{CAV Evaluation in Mixed Traffic Conditions}
Three main approaches have been used to assess the benefits of CAVs: 
\begin{enumerate*}[label=\arabic*)]
\item analytical study,
\item simulation evaluation, and
\item field test with equipped vehicles.
\end{enumerate*}
On-road testing provides the utmost degree of realism with equipped automated driving systems (ADS) and real-world traffic environment. However, the safety and efficiency issues for testing CAV on public roads have been the major concern, especially after several severe CAV-involved accidents in recent years. Due to safety, technological, and budgetary limitations, the scale of a CAV field test at current stage tends to be small (e.g., with a handful of CAVs). As a result, the conclusions from these small-scale field tests may not be reliably generalized to a traffic flow level. Furthermore, it was estimated by \citeauthor{Kalra2016} that billions of kilometers of road test would be required to achieve the desired level of confidence in terms of safety of an ADS \cite{Kalra2016}. Thus, analytical and simulation approaches serve as two primary methods for evaluating traffic flow impact of CAVs. 

The majority of the analytical models is based on macroscopic traffic flow models and may experience difficulty in faithfully capturing the complex phenomena in transportation networks, such as lane drop. 
\citeauthor{smith2015benefits} proposed an analytical framework for assessing the benefits of CAV operations \cite{smith2015benefits}. The results indicated that CAVs improved network mobility performance, even with low MRP and no managed lane policies. Throughput without managed lanes increased by 4\%, 10\%, and 16\% at the MPR of 10\%, 20\%, and 30\%, respectively. It was also discovered that the managed lane policy facilitated homogeneous CAV traffic flow leading to more consistent and stable network outputs. 
An analytical model for determining the optimal managed lane strategy was proposed in \cite{hussain2016freeway}, where the maximum system throughput in a mixed traffic condition could be calculated under the assumption of random vehicle distribution on a freeway facility. Three types of headways (i.e., conservative, neutral, and aggressive) were used in the model.
\textcolor{r1}{
\citeauthor{wang2017Comparing} proposed a second-order traffic flow model for mixed traffic streams with HVs and AVs. The authors found that the second-order model consistently outperformed the first-order one in terms of the accuracy of traffic density when the variability of the penetration rate increases \cite{wang2017Comparing} . 
}


At the corridor level, a capacity of 4,250 vphpl (vehicle per hour per lane) was observed in \cite{van2006impact} on a 6-km highway segment with uniformly distributed ramps under full market penetration of CAVs. The study by \citeauthor{shladover2012impacts} observed a pipeline capacity of 3,600 at 90\% MPR of CAVs \cite{shladover2012impacts}, where the pipeline capacity refers to the throughput observed on a single-lane roadway without any interference of lateral movements \cite{Hall1997capaciy}.
\citeauthor{Arnaout2014} evaluated CAVs under moderate, saturated, and over-saturated demand levels on a hypothetical 4-lane highway under different market penetrations. They found that 9,400 vehicles could be served within an hour when the CAV MPR reached 40\% \cite{Arnaout2014}. \citeauthor{songchitruksa2016} assessed the network performance with CAVs on the 26-mile I-30 freeway in Dallas, TX and found the highest throughput being 4,400 vph at 50\% MPR \cite{songchitruksa2016} among four MPR scenarios (i.e., 10\%, 30\%,  50\%, and 70\%). Another study \cite{lee2014mobility} also revealed that the mobility benefits of CAV emerged at 30\% MPR. 

\citeauthor{Liu2019Freeway} investigated the benefits of alleviating freeway merge bottleneck and compared the performance of CACC with ACC under full market penetration. The results showed that CACC yielded a 50\% reduction in fuel consumption  (as estimated with the EPA MOVES model) while increasing corridor capacity by 49\%, compared to the ACC scenario \cite{Liu2019Freeway}.
With a subsequent test on an 18-km segment of SR-99, the research team found that deploying vehicle awareness device (VAD)-equipped vehicles along with managed lane strategies was helpful in improving corridor-level traffic flow under low and medium CAV market penetrations \cite{Liu2018Modeling}. 
\textcolor{r1}{Besides MOVES, comprehensive modal emission model (CMEM) \cite{an1997development}, VT-Micro \cite{Rakha2004}, the Future Automotive Systems Technology Simulator (FASTSim)\cite{brooker2015fastsim} are among the commonly used vehicle emission models in quantifying potential environmental impact of deploying CAVs.} 


\textcolor{r1}{The potential impact of the short following time headway of CAVs on HVs has also been studied in previous studies. Among them, the KONVOI project found that the carry-over effect for CACC drivers in manual driving after the disengagement of the CACC system \cite{casey1992changes}. In recent years, driving simulator has been employed to study the behavioral adaptation of human drivers operating in the vicinity of CAVs. \citeauthor{nowakowski2011cooperative} found that test participants are likely to drive under a shorter following distance in the presence of CACC platoons in the adjacent lane  \cite{nowakowski2011cooperative}. A similar study was conducted by \citeauthor{GOUY2014264} to investigate the behavioral adaptation of human drivers along a CACC platoon, in which two CACC platoon configurations were tested: 1) a 10-truck platoon with 0.3-s intra-platoon headway and 2) a 3-truck platoon with 1.4-s intra-platoon headway. It was found that a smaller average HV headway was observed in the first scenario, under which participants spent more time under a 1-s headway. Although such short headway was generally deemed unsafe in previous studies (e.g., \cite{FAIRCLOUGH1997387}).
}

\subsection{CAVs and Managed Lanes}

Managed lanes have been in practice over the years to improve target operation objectives, such as
\begin{enumerate*}[label=\arabic*)]
\item promoting the adaptation of environmentally-friendly vehicles by offering priority usage to specific travel lanes (e.g., the California Clean Air Vehicle Decal \cite{shewmake2014hybrid}),
\item encouraging car-pooling by adopting high-occupancy vehicle (HOV) lanes \cite{chang2008review}, and
\item performing active traffic management with the aid of high-occupancy toll (HOT) lanes \cite{gomez2018toll}.
\end{enumerate*}
a CAV lane is one variant of managed lane strategies that provides exclusive lane use privilege to CAVs. Although managed lane strategies have been widely applied to highway operation with successful cases, due to distinctive operational characteristics of CAVs, knowledge learned from a conventional managed lane may not be directly transferable to the implementation of a CAV lane.

The provision of a CAV-managed lane has two primary reasons. First, CAV-managed lanes can incentivize the adaptation of CAVs by offering priority usage to managed lanes, which typically provides better and more reliable travel because of active traffic management.
More importantly and unique to CAVs, CAV-managed lanes can provide accommodations for the underlying operational characteristics of CAVs.
\textcolor{r1}{
A CAV is able to operate at a much closer headway than a human driver with the assistance of V2V wireless communication and the automated driving system (ADS) \cite{Nowakowski2011,shladover2018using}. Hence, the necessary condition for realizing such a short following headway is the availability of the vehicle driving information of at least one of the predecessors on the same lane, i.e., through a CAV following another CAV. Otherwise, the string stability of CAVs cannot be guaranteed \cite{swaroop1994comparision}, and the lack of thereof is termed as CAV degradation \cite{Wang2019stability}, which could potentially be a major hurdle for CAVs operating in mixed traffic. A numerical example by \citeauthor{Wang2019stability} has showed that the current technological maturity of CACC contributed negatively to the stability of heterogeneous flow \cite{Wang2019stability}.
}

To mitigate CAV degradation, ad hoc coordination, local coordination, and global coordination are the three major strategies that outline the organization of CAV platoons \cite{shladover2015cooperative}.
Ad hoc coordination assumes that CAVs arrive in random sequence and do not actively seek clustering opportunities in a traffic stream. By extension, the probability of driving around other CAVs is highly correlated to MPR. On the contrary, CAVs actively identify and approach an existing CAV cluster (or other free-agent CAVs) to form a new cluster through local coordination, regardless of CAV MPRs. 
Finally, global coordination (a.k.a. end-to-end platooning) requires a high-level route planning and extensive communication to coordinate vehicles traveling with the same origin-destination pair even before the CAVs entering highway sections \cite{larson2013coordinated}. 

To successfully form and maintain platoons, accurate and cost-effective localization of CAVs in a dynamic traffic environment remains one of the biggest challenges, especially for local coordination \cite{shladover2015cooperative,kuutti2018survey}. In the presence of a CAV-managed lane, a higher concentration of CAVs facilitates local coordination with much less stringent requirements on the accuracy of vehicle localization.
In addition, the CAV-managed lane strategy aligns well with the three-stage deployment roadmap considering market diffusion and technological maturation for CAVs \cite{ZhongDis}. In the first stage, the adoption of CAVs is incentivized by allowing the use of the managed lane free of charge. At this stage, the following headway of CAVs on the managed lane may be comparable to that has been observed from HVs for safety reasons in a mixed traffic condition. In the second stage, a shorter following headway for CAVs could be implemented to further increase the carrying capacity of the managed lane when the demand of CAVs along with the familiarity of road users to CAVs increases. In the third stage, when the CAVs reach a critical level of MPR, high-performance driving enabled by the CAVs can be achieved due to  homogeneous CAV traffic flow on the managed lane.



To assess the impact of CAV-managed lane strategies, \citeauthor{zhang2018operational} compared the performance of a managed lane and general propose lanes (GPL) based on average speed, throughput, and travel time \cite{zhang2018operational}. The results indicated that the speed improvement in the managed lane was significant compared to that of GPLs. With 20\% MPR, the latent demand (the demand that cannot enter the simulation network due to congestion) decreased to zero. 
Inspired by the fluid approximation of traffic, \citeauthor{wright2018dynamic} proposed an algorithm for simulating the weaving activity at the interface of a managed lane and the adjacent GPL at a macroscopic scale \cite{wright2018dynamic}. 
\citeauthor{chen2016optimal} proposed a time-dependent deployment framework that was formulated with a network equilibrium model and a diffusion model. With the constraint of a given set of candidate lanes which corresponds to the field condition, the social cost was minimized with the consideration of different MPR levels \cite{chen2016optimal}. \citeauthor{Zhong2019Effectiveness} studied four managed lane strategies and compared the benefits for GPL and managed lane users in terms of mobility, safety, emission, and equity \cite{Zhong2019Effectiveness}. In freeway settings, the authors recommended a 30\% minimal MRP for deploying a CAV-managed lane to avoid lane use imbalance that could degrade the performance \cite{Zhong2018Assess, Zhong2018Simulation}.

\citeauthor{qom2016evaluation} proposed a multi-resolution framework to study the mobility impact of CAV lanes. Traffic flow-based static traffic assignment and the meso-scopic simulation-based dynamic traffic assignment were adapted in the bi-level framework. The former yielded the MPR-based trends, whereas the latter refined the trend based on traffic congestion. The results indicated that it was not beneficial to provide toll incentive for CAVs at lower MPR due to the marginal increase in highway capacity \cite{qom2016evaluation}.
\citeauthor{ghiasi2017mixed} proposed an analytical capacity model for mixed traffic \cite{ghiasi2017mixed}. The model relied on the Markov chain representation of the spatial distribution of heterogeneous and stochastic headway.  With the sufficient and necessary condition of capacity increase proven, the authors emphasized the importance of quantitative analysis of the actual headway stetting. 

The introduction of a CAV lane to a signalized corridor was reported in \cite{zhong2017evaluations}. Two configurations of a CAV lane, along with other managed lanes, were evaluated. Due to the turning nature of an arterial, buffer zones were implemented where HVs are allowed to temporarily use the CAV lane for turning movements.  \citeauthor{papadoulis2019evaluating} evaluated the safety impact of CAVs using the Surrogate Safety Assessment Model (SSAM) \cite{papadoulis2019evaluating}. The time to collision (TTC) and the post encroachment time (PET) were adapted with safety thresholds of 1.5 s and 5 s, respectively. They observed substantial safety benefits in terms of reduction in traffic conflicts: 12-47\% at 25\% MPR to 90-94 \% at 100\% MPR. In \cite{Zhong2020Influence}, TTC was also used to assess the safety conditions for HVs when CAV local clustering strategy was employed. 
\textcolor{r2}{
\citeauthor{ali2018connectivity} found that drivers with advanced traffic information enabled by connectivity tend to wait longer and maintained a larger space on mandatory lane change (the communication delay for lane merging assistance was unnoticeable when it was less than 1.5 s). Post-encroachment time (PET) analysis also indicated improved travel safety from CAV implementation 
\cite{ali2018connectivity}. 
}

\subsection{Summary}
The vast majority of previous studies evaluated the benefits of CAVs at an aggregated level with the emphasis of overall traffic improvement. Analytic models are in macroscopic nature under overly ideal conditions, and they have difficulty in factoring in the stochastic nature of human drivers in a mixed traffic environment. CAV-managed lane strategy could be instrumental in the near-term deployment of CAVs, but it is still an under-explored area, despite its increasing recognition. 



\section{Evaluation Framework and Experiment Design}
\label{sect:methodology}
This study focuses on analyzing mixed traffic flow characteristics at a corridor level considering different CAV MRPs and managed lane strategies. In this section, the integrated simulation test bed, transportation network, and simulation scenarios are discussed in detail.


\subsection{CAV Behavior Model}
The PTV Vissim \cite{ptv2018}, a \textcolor{r1}{commercial-off-the-shelf} microscopic simulation package, is chosen for the evaluation. Vissim has been widely adapted by transportation practitioners and researchers, owing to its high-fidelity simulation mechanism and flexible modules. Although compared to other open-source traffic simulators (e.g., SUMO), one reservation for Vissim being a commercial software is its close-sourced nature. 
As shown in Table \ref{table:vehControl}, a calibrated Wiedemann car-following model and the Enhanced Intelligent Driver Model (E-IDM) \cite{kesting2010enhanced} were used to model HVs and CAVs, respectively. 
The intelligent driver model (IDM) and its variants have been used to design the ACC/CACC controller that resembles human-like car-following behaviors \cite{Wang2019Benefits,Kesting2008adaptive,Talebpour2016Modeling,Spiliopoulou2017Exploitation,gueriau2016assess}.
As an improved iteration of the collision-free IDM \cite{Treiber2000}, the E-IDM deals with CAV longitudinal maneuver. The behavior model of the E-IDM is expressed in Eq. \ref{eq:eidm}, \ref{eq: minDistCal}, and \ref{eq: cahCal}.

\begin{table}[h]
\center
\caption{\textcolor{r1}{Differences between HVs and CAVs in the simulation model}}
\begin{tabular}{l|llc}
\hline
Vehicle Type & Longitudinal Control & \textcolor{r1}{DTG} & \textcolor{r1}{Stochasticity} \\ \hline
HV & Weidemann 99  & \textcolor{r1}{1.4 s} & \textcolor{r1}{Y} \\
CAV & E-IDM  & \textcolor{r1}{0.6, 1.2 s} & \textcolor{r1}{N} \\
\hline
\end{tabular}
\label{table:vehControl}
\end{table}


\begin{equation}
\ddot{x}=\begin{cases}
a[1-(\frac{\dot{x}}{\dot{x_{des}}})^{\delta }- (\frac{s^{*}(\dot{x}, \dot{x}_{lead})}{s_{0}})] &  \qquad\text{if }  \ddot{x}_{IDM} \geq \ddot{x}_{CAH} \\ 
 (1-c)\ddot{x}_{IDM} + c[\ddot{x}_{CAH} + b \cdot tanh ( \frac{\ddot{x}_{IDM} - \ddot{x}_{CAH}}{b})] &  \qquad\text{otherwise} 
\end{cases}
\label{eq:eidm}
\end{equation}
\begin{equation}
s^{*}(\dot{x}, \dot{x}_{lead}) = s_{0} + \dot{x}T + \frac{\dot{x}(\dot{x} - \dot{x}_{lead})}{2\sqrt{ab}} 
\label{eq: minDistCal}
\end{equation}
\begin{equation}
\ddot{x}_{CAH}=
\begin{cases}
\frac{\dot{x}^{2} \cdot \min\{\ddot{x}_{lead}, \ddot{x}\}}{\dot{x}_{lead}^{2}-2x \cdot \min\{\ddot{x}_{lead}, \ddot{x}\}} &  \qquad\text{if }  
\dot{x}_{lead} (\dot{x} - \dot{x}_{lead}) \leq  -2x \cdot \min\{\ddot{x}_{lead}, \ddot{x}\}  \\ 
\min\{\ddot{x}_{lead}, \ddot{x}\} - \frac{(\dot{x}-\dot{x}_{lead})^{2} \Theta (\dot{x}- \dot{x}_{lead})}{2x}  &  \qquad\text {otherwise}
\end{cases} 
\label{eq: cahCal}
\end{equation}
where, $a$ is the maximum acceleration; $b$ is the desired deceleration; $c$ is the coolness factor; $\delta$ is the free acceleration exponent; $\dot{x}$ is the current speed of the subject vehicle;  $\dot{x}_{des}$ is the desired speed,  $\dot{x}_{lead}$ is the speed of the lead vehicle; $s_{0}$ is the minimal distance; $\ddot{x}$ is the acceleration of the subject vehicle; $\ddot{x}_{lead}$ is the acceleration of the lead vehicle; $\ddot{x}_{IDM}$ is the acceleration calculated by the original IDM model \cite{Treiber2000}. The minimal distance can be calculated as $s^{*}(\dot{x}, \dot{x}_{lead})$ where  $T$ is the desired time gap; and $\ddot{x}_{CAH}$ is the acceleration calculated by the constant-acceleration heuristic (CAH) component, where $\Theta$ is the Heaviside step function \textcolor{r1}{that is used to eliminate the negative approaching rate of subject vehicle \cite{kesting2010enhanced}}. 

In this study, the E-IDM model is selected as the longitudinal control for the CAVs. 
\textcolor{r1}{
Although without built-in multi-anticipative car-following function, as the literature shows, E-IDM is still a good simple car-following model for CAVs, as the stochastic nature of human driving is removed (i.e., automation property), and the acceleration of the preceding vehicle is taken into account in the driving model (i.e., connectivity property). }
As shown in Table \ref{table: parameters}, all the parameters remain the same as those originally specified in \cite{kesting2010enhanced}, with the exception of the desired time gap (DTG), which is defined with two values: 0.6 s and 1.2 s. The former DTG is used when the communication between a preceding CAV and the subject CAV is successful, whereas the latter one is in effect when the communication failure occurs. The updating frequency for the E-IDM model in Vissim is 10 Hz. The density of CAVs which is used to calculate the communication activity is updated at a 2-Hz frequency to reflect the traffic dynamic. Each transmission is assumed to have up to five attempts (four re-transmission).  At least one successful attempt is required for a transmission to be considered successful, upon which the DTG is determined.  
\begin{table}[h]
\centering
\caption{E-IDM Vehicle Control Parameters} 
\begin{tabular}{c|cccccccccc}
\hline 
Parameter & $T_{intra}$ & $T_{inter}$ & $s_{0}$ & $a$ & $b$ & $c$ & $\theta$ & $\dot{x}_{des}$ \\ \hline
value & 0.6 s & 1.2 s & 1$m$ & 2$m/s^{2}$ & 2$m/s^{2}$ & 0.99 & 4 & 105 $km/h$  \\
\hline 
\end{tabular}
\label{table: parameters}
\end{table}

\textcolor{r1}{To implement these two car following models in Vissim, the subset of the human driving behavior is realized by adjusting car-following parameters of the Wiedemann car-following model, which is relatively straightforward.
The E-IDM, on the other hand, is implemented via the external driver model application programming interface (API) and connected with Vissim through a dynamic link library (DLL). The DLL is invoked in each simulation time step such that the default car-following behavior will be overwritten for a specified vehicle type. The DSRC wireless communication module, discussed later in Section \ref{sect: dsrc}, is also implemented in the API to achieve a dynamic response based on prevailing traffic conditions.}

\textcolor{r2}{
One of the most prominent features in CAV behavior modeling is the short time headway during car-following, which is manifested by several key differences between a CAV and a HV. First, the stochasticity of the CAVs is significantly lower than that of human drivers. This is enabled by the on-board sensors that are able to continuously and accurately perceive the surrounding environment. However, the stochaticity cannot be completely eliminated due to sensor noise and communication delay/error.  Second, a CAV has minimal reaction time due to its algorithmic decision making process and computational power. Past studies have already identified that the impact of the reaction time of human drivers in various traffic phenomena, including capacity drop \cite{calvert2018capacity} and flow stability \cite{Talebpour2016Modeling}, whereas driving simulation tests revealed that the information augmented by connectivity could decrease the reaction time for drivers \citep{sharma2019estimating}. }

\textcolor{r2}{
In addition, human factor plays a crucial role in the resumption of control of a CAV when an ADS exits its operational domain (e.g., high risk of collision, sensor failure, communication interference). 
Quantitative evidence regarding the transition of control from traffic psychology or human-machine interactions is still limited \cite{calvert2020generic}, though few  frameworks have been proposed to simulate human behavior endogenously \cite{venLint2018generic, Saifuzzaman2017understanding}. For example, the prospect theory was used to model the risk and human perception \cite{hamdar2015behavioral,sharma2019modelling}. The Risk Allostasis Theory \cite{fuller2011driver} was adopted for modeling relationship between cognitive processing of information and physical performance. The Task Capacity Interface \cite{saifuzzaman2015revisiting} was employed by \citeauthor{saifuzzaman2015revisiting} for quantifying situational awareness of a driver.}

\textcolor{r2}{\citeauthor{calvert2020generic} developed a framework that encompasses the driving task demand and driver task saturation \cite{calvert2020generic}. The framework's main goal is to assess the performance impact during the transition of control for AVs. The total task demand, situational awareness, and reaction time during the transition of control from AVs was explored. The framework showed promising capability in capturing the interactive effects of humans with lower level AVs. However, empirical evidence is still needed to relax the assumptions used in the framework from not only the cognitive point of view, but also from vehicle dynamics and inter-vehicle interactions.
}

\textcolor{r2}{
Another human factor is driver compliance to the ADS. Since in lower or medium level of automation, the driver is ultimately responsible for his or her vehicle, which means overwriting, when deemed necessary, is possible by the human driver. This control authority, in extreme cases, could cancel out the benefits promised by the CAV technologies. In a recent study \citep{sharma2019modelling}, \citeauthor{sharma2019modelling} employed the prospect theory to model driver decision-making mechanisms including irrational ones, and captured the negative relationship between headway and compliance decision by a driver.
}


\textcolor{r2}{
In this study, we represent the differences of a CAV and a HV with different desired time headways through separate car following models, with the following assumptions made for CAVs: 1) no error for the on-board sensors and the vehicle controller, i.e., perfect perception; 2) no human factor modeling pertaining to the transition of authority; and 3) no behavior adaptation for CAVs for non-CAV drivers.
}

\subsection{Wireless Communication Model}\label{sect: dsrc}
In an early study, we implemented a packet-level communication module through Vissim API  \cite{Zhong2018Simulation}. Similar adaptations for the model were also found in previous studies \cite{songchitruksa2016,Zhong2018Simulation,Bieka2019assessing}. The analytical model \cite{killat2007enabling} was developed from ns-2, an empirical packet-level network simulator that returns the probability of one-hop broadcast reception of basic safety message (BSM) under IEEE 802.11p, an approved amendment tailored to wireless access in vehicular environment (WAVE) in the 802.11 family protocol. The model uses the concept of communication density level, a metric representing channel load in vehicular communication in the form of the sensible transmission per unit of time and per unit of the road \cite{Jiang2007communication}.
\textcolor{r1}{The data reception rate is determined jointly by communication density level and transmission power. An illustration for the reception probability is shown in \ref{app:coefficent}.} Note that this communication model only pertains to the physical layer of the DSRC communication (e.g., no MAC layer delay).

\begin{equation}
\label{eq:dsrcMod}
\begin{aligned}
P_{r}(x, \delta , \varphi , f)= & e^{^{-3(x/\varphi )^{2}}}\Big(1+ \sum_{i=1}^{4}h_{i}(\xi ,\varphi )(\frac{x}{\varphi})^{i}\Big) \\
\xi = & \delta  \cdot \varphi \cdot f
\end{aligned}
\end{equation}

%

where, ${h_{i}(\xi ,\varphi)}$ is the two-dimensional polynomial of fourth-degree for all curving fitting parameters \cite{Killat2009}, which is also shown in \ref{app:coefficent}; $\xi$ is communication density, events/s/km; and $\varphi$ is the transmission power, m;  $\delta $ is vehicle per kilometer that periodically broadcast messages, veh/km; and $f$ is transmission rate, Hz.

\subsection{Transportation Network}
\begin{figure}[h]
	\centering
	\includegraphics[width=0.8\textwidth]{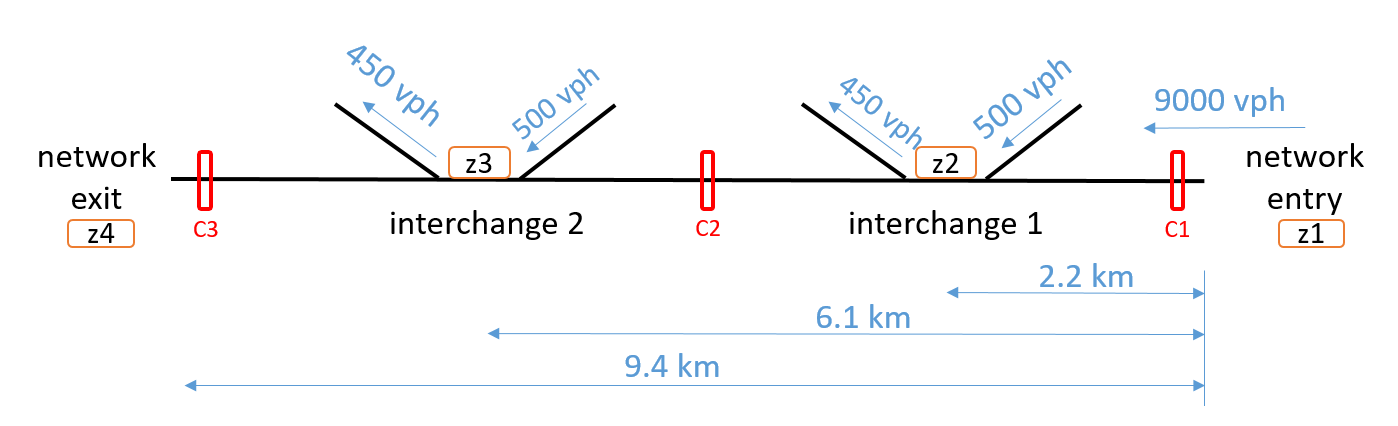}   
	\caption{Network geometry and demand} 
	\label{fig:networkGeometryDemand}
\end{figure}

A 9.3-km 4-lane hypothetical network was constructed in Vissim with two interchanges located at mile marker 2 (km) and 6 (km) respectively. An abstract geometry of the network along with vehicle demand of the origins and destination is shown in Figure \ref{fig:networkGeometryDemand}. The primary reason for using a simply synthetic network is to limit variables for the simulation. Note that the driving behavior parameters for the Wiedemann car-following model (for HVs) is the same as in previous studies \cite{STOLT4,Li2019High,Zhong2018Assess,Zhong2020Influence}, which represents a subset of the calibrated driving behavior in the I-66 segment in northern Virginia. The demand originated on the mainline is deliberately set higher than usual to create a congested network.
\textcolor{r1}{The speed limit for the mainline of the network is set as 120 $km/h$.} Three data collectors are placed at ``C1'', ``C2'', and ``C3'' locations.

\subsection{Managed Lane Scenarios}
Three cases of CAV lanes, as shown in Table \ref{table:mlCase}, are implemented in the network:

\begin{itemize}
\item{\textbf{No managed lane (NML)}: This scenario serves as the base condition of the study. There is no priority lane use for CAVs, and they are mixed with HVs throughout the network;}
\item{\textbf{One CAV lane (CAV-1)}: In this strategy, one CAV lane is implemented in the left-most lane (the fourth lane from the right);}
\item{\textbf{Two CAV lanes (CAV-2)}: An additional CAV lane is added to the CAV-1 case, making two CAV lanes available at the leftmost and the second leftmost lane in the roadway segment. It aims to investigate the duel managed lane configuration.}
\end{itemize}

As revealed in \textcolor{r1}{previous studies \cite{ZhongDis,zhang2018operational,xiao2019traffic, NCHRP}, a managed lane may have a detrimental effect on traffic performance if implemented prematurely, i.e., usually with an MPR less than 30\%. Therefore in this study, we set CAV MPRs for ``CAV-1'' to start from 30\%.  With the same logic, the ``CAV-2'' cases start with 40\% to cover certain transition MPR, since the linear extrapolation may not hold.}

\begin{table}[!hbtp]
\centering
\renewcommand{\arraystretch}{1.2} 
\caption{Managed Lane Evaluation Plan} 
\begin{tabular}{l|ccc}
Policy & No Managed Lane & Managed Lane \#1 & Managed Lane \#2 \\ \hline
\textbf{ID} & \textbf{NML} & \textbf{CAV-1} & \textbf{CAV-2} \\ \hline
\textbf{1st Lane} & HV + CAV& HV + CAV& HV + CAV\\
\textbf{2nd Lane} & HV + CAV& HV + CAV& HV + CAV\\
\textbf{3rd Lane} & HV + CAV& HV + CAV& CAV\\
\textbf{4th Lane} & HV + CAV& CAV& CAV\\
\textbf{MPR} & 0\% - 100\% & 30\% - 100\% & 40\% - 100\%\\
\hline
\end{tabular}
\label{table:mlCase}
\end{table}

\section{Results and Analysis}
\label{sect:result}
Five replications are run for each combination of managed lane policies and MPRs. Aggregated data are collected at 5-min intervals, and the raw data are collected at each simulation time step. The analysis is performed on four aspects:
\begin{enumerate*}[label=\arabic*)]
\item traffic flow characteristics, 
\item headway distribution,
\item fuel consumption,
\item wireless communication, and 
\item overall network performance.
\end{enumerate*}

\begin{figure*}[]
	\centering
	\frame{\includegraphics[width=.95\textwidth]{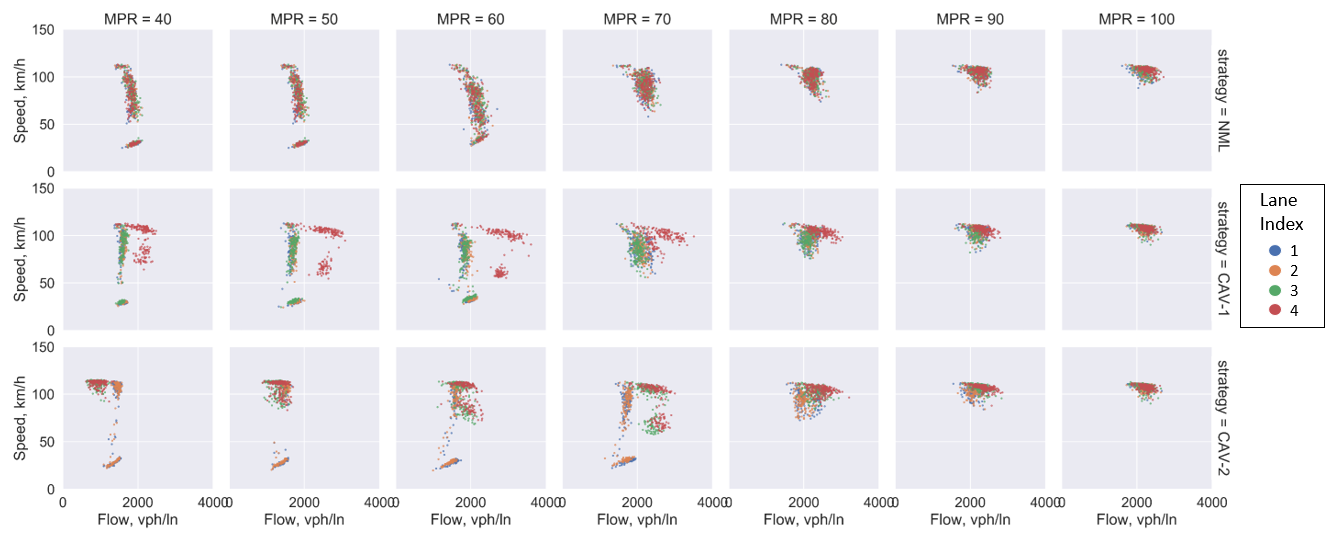}   }
	\caption{Speed-flow curves} 
	\label{fig:spdFlowCurve}
\end{figure*}

\subsection{Traffic Flow Characteristic} 
Figure \ref{fig:spdFlowCurve} exhibits the speed-flow characteristics of the simulation scenarios having 40\% MPR and above. The plot is color-coded by travel lanes with index ``1'' representing the rightmost lane, and ``4'' the leftmost.  The speed-flow diagram is comprised of a stable region and a unstable (congested) region, separated by the optimum (maximum) flow. Several distinctive patterns can be observed.  First, regardless of the managed lane strategy, the sample points become more concentrated as the MPR increases, with the disappearance of the congested region typically found in the lower speed region. Second, the CAV lane has a distinct pattern compared to the GPLs. Such pattern is most apparent in CAV-1, where the traffic samples on the leftmost lane (CAV lane) shift to the right along the flow axis. The congested region disappears when MPR reaches 70\% in the CAV-1 case for all of the lanes. The improvement for the GPs is due to a higher carrying capacity of the CAV lane, which results in less traffic on the GPLs. The homogeneity of the CAV traffic is the primary factor in realizing the mobility benefit of CAVs: in NML cases, the sample points from difference lanes are evenly distributed, in contrast to managed lane cases. For the CAV-2 case, the separation of the CAV lanes (leftmost and the second leftmost) started to show at 70\% MPR. At full penetration (100\%), the traffic patterns are very similar, as the managed lane becomes irrelevant.


\subsection{Headway Distribution}
The simulation collects raw data from the data collector, an equivalent of real-world detectors (e.g., loop detectors, video cameras, microwave sensors). By analyzing the high-resolution raw data (collected every 0.1 s), the headway distribution in CAV lanes can be obtained. Recall that the collectors are placed in three sections of the roadway segment, as shown in Figure \ref{fig:networkGeometryDemand}.

\begin{figure*}[]
	\centering
	\includegraphics[width=\columnwidth]{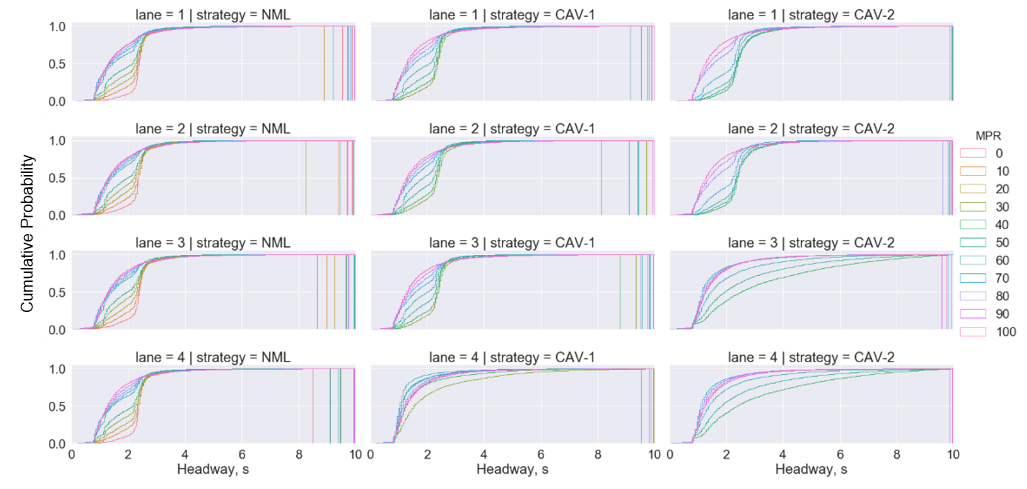}   
	\caption{CDF for headway distribution among travel lanes} 
	\label{fig:cdfHwLane}
\end{figure*}

The cumulative probability function (CDF) curves are displayed in Figure \ref{fig:cdfHwLane}. The vertical lines in the figure are the headways when 100\% cumulative probability is reached. The slope of the CDF indicates the level of concentration of the samples within a distribution. 
In NML cases, two types of tipping points exist: the one at lower headway resulted from a high MPR and the one with higher headway observed at a low MPR (below 40\%).  For CAV-1, the pattern for CDF at 30\% and 40\% is transformed to the pattern observed at high MPRs. With 2 CAV lanes, the CDF increases gradually in the mid-range MPR (40\% to 60\%) because of under-saturation on the CAV lanes, as illustrated in the CDF on the 3rd and 4th lanes. Such under-saturation situation is alleviated when the MPR reaches 70\%.
A similar pattern in CDFs is observed at a high MPR range (i.e., 80\% to 90\%) regardless of the managed lane strategies, indicating a high concentration of samples with headway above 1 s.


\begin{figure} [!h]
	\centering
	\frame{\includegraphics[width=0.7\columnwidth]{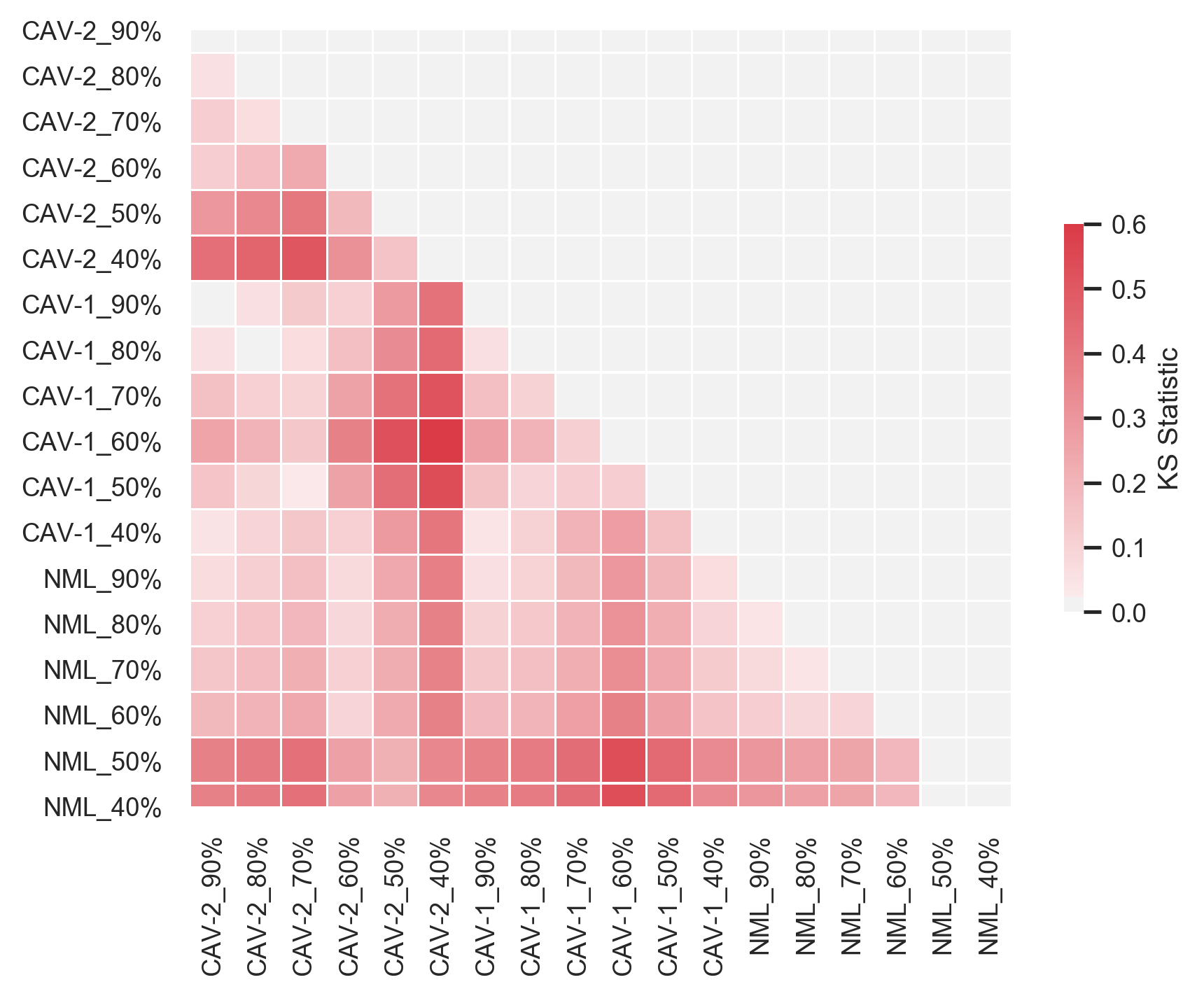}}   
	\caption{K-S statistics for CDF comparison} 
	\label{fig:ksStat}
\end{figure}

Two-sample Kolmogorov-Smirnov (K-S) test is adopted to analyze the CDFs to check whether two random samples are from the same population \cite{goodman1954kolmogorov}. It is a non-parametric test where no assumption is made regarding the distribution of the variables \cite{young1977proof}. The null hypothesis ($H_0$) of the two-sample K-S test is that the two sample sets are from the same continuous distribution.  
Nearly all the CDFs in the pairwise comparison rejects the null hypothesis with a low p-value at the 0.05 significance level, with the exception of the comparison of 40\% and 50\% in NML. Figure \ref{fig:ksStat} is a heatmap that shows the pairwise K-S statistic that represents the supremum of the two tested empirical CDFs. The denser the color, the higher the difference in cumulative probability between two comparing scenarios.

\begin{figure} [h]
	\centering
	\frame{\includegraphics[width=0.8\columnwidth]{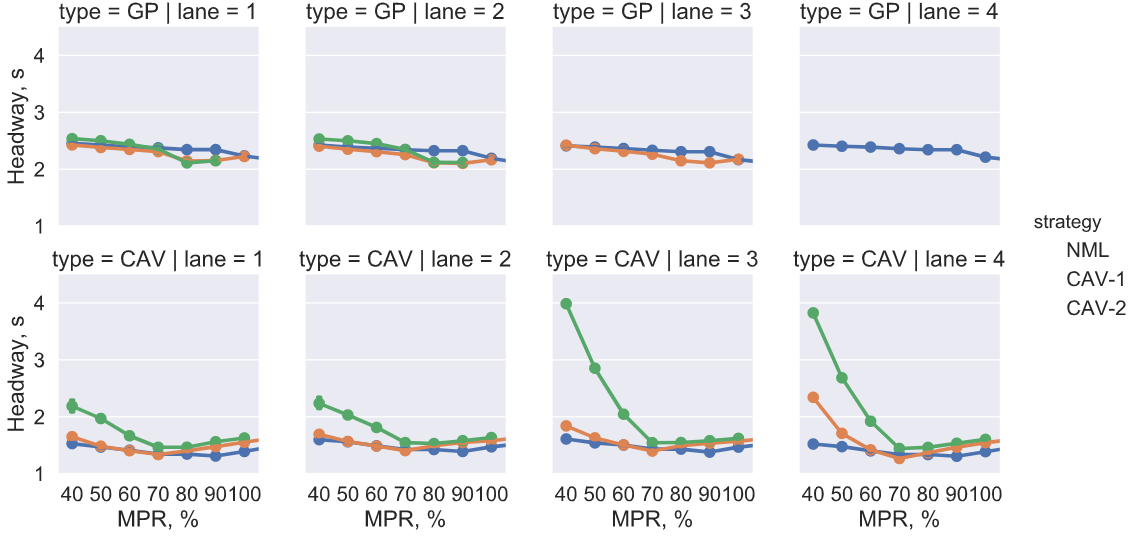}}
	\caption{Average headway} 
	\label{fig:avgHw}
\end{figure}

The average headway for HVs and CAVs in every travel lane is shown in Figure \ref{fig:avgHw}. The row represents the vehicle types, whereas the column represents the travel lane. Recall that the 4th lane is the leftmost lane. For HVs, their averaged headway decreases as the MPR increases in CAV-1 and CAV-2 cases. While the headway also decreases in the NML case, it is at a lesser rate. When it comes to CAVs, the decreasing rate in CAV-2 is greater than that in CAV-1 or NML. The mean headway is around 4 s in CAV-2 case when the MPR is low or in mid-range due to low lane utilization in the CAV lanes. The average headway in CAV-2 case reaches a comparable level to its counterparts at 70\% MPR, which is the deflection point. The lowest mean headway achieved among all scenarios is observed at 70\% MPR in CAV-1 case for CAVs, which corresponds to the maximum capacity with all other factors being equal. Lastly, the headway trend for CAVs remains a similar pattern across four travel lanes in the NML case, since CAVs are uniformly distributed across all lanes.

\begin{figure*}[!h]
\begin{minipage}[h]{0.5\textwidth}
\centering
\subfloat[40\% MPR]{\label{main:a}\includegraphics[scale=.6]{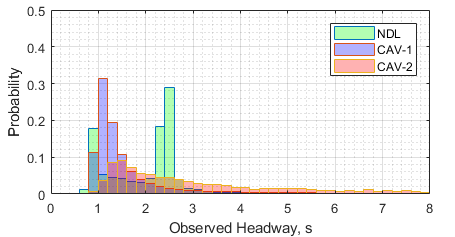}}
\end{minipage}
\begin{minipage}[h]{0.5\textwidth}
\centering
\subfloat[50\% MPR]{\label{main:b}\includegraphics[scale=.6]{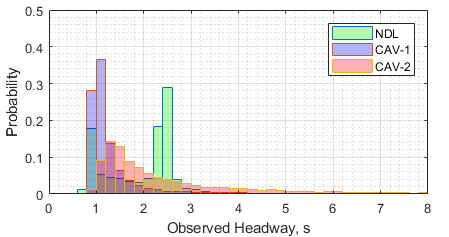}}
\end{minipage} \par
\begin{minipage}[h]{0.5\textwidth}
\centering
\subfloat[60\% MPR]{\label{main:c}\includegraphics[scale=.6]{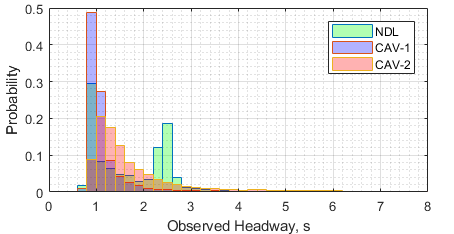}}
\end{minipage}\
\begin{minipage}[h]{0.5\textwidth}
\centering
\subfloat[70\% MPR]{\label{main:d}\includegraphics[scale=.6]{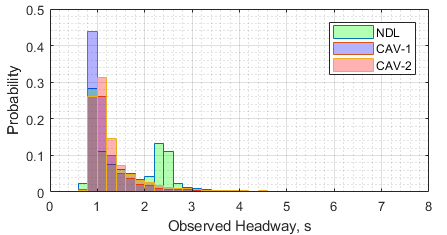}}
\end{minipage}
\caption{Headway distributions in the leftmost lane}
\label{fig:hwDistribution}
\end{figure*}

Figure \ref{fig:hwDistribution} shows the comparison of headway distributions in the leftmost lane among three managed lane scenarios under different MPRs. In the 40\% to 70\% MPR range, it is shown that implementing a managed lane for CAVs clearly shifts the distribution to the left-hand side, which represents smaller headways. The distributions of headway collected for either CAV-1 or CAV-2 become ``narrower'' (with less standard deviation), as the MPR increases from 40\% to 70\%. The highest bin of the histogram for both CAV-1 and CAV-2 cases is 1 s - 1.2 s when the MPR is below 50\%. When the MPR is higher than 50\%, the highest bin of the histogram shifts to 0.8 s - 1 s. In comparison, the NML case does not exhibit such a concentration pattern as the MPR increases. The result indicates that a homogeneous traffic flow comprised of only CAVs is able to realize the short-headway benefits from deploying CAVs.

\subsection{\textcolor{add}{Fuel Consumption}}
\textcolor{add}{
The VT-Micro model \cite{Rakha2004}, an individual vehicle and operation-level emission model, is adopted to calculate the instantaneous fuel consumption rate. The inputs for the VT-Micro model are instantaneous vehicle speed and acceleration, and the output is the second-by-second fuel consumption rate, as shown in Equation \ref{eq:vtMicro}, where $\dot{x}$ is the instantaneous speed, $\ddot{x}$ is instantaneous acceleration, $L^{e}_{i,j}$ and $M^{e}_{i,j}$ are regression coefficients for emission and fuel consumption at speed power $i$ and acceleration power $j$, respectively. 
\begin{equation} 
\label{eq:vtMicro}
f(\dot{x}, \ddot{x})=
\left\{\begin{matrix}
 exp\Big(\sum^{3}_{i=0}\sum^{3}_{j=0}(L^{e}_{i,j} \cdot \dot{x}^{i} \cdot \ddot{x}^{j})\Big)
 & \text{for } \ddot{x} \geq 0\\ 
 exp\Big(\sum^{3}_{i=0}\sum^{3}_{j=0}(M^{e}_{i,j} \cdot \dot{x}^{i} \cdot \ddot{x}^{j})\Big) 
& \text{for } \ddot{x} < 0 
\end{matrix}\right.
\end{equation}
}

\begin{figure*}[h]
	\centering
	\includegraphics[width=1\columnwidth]{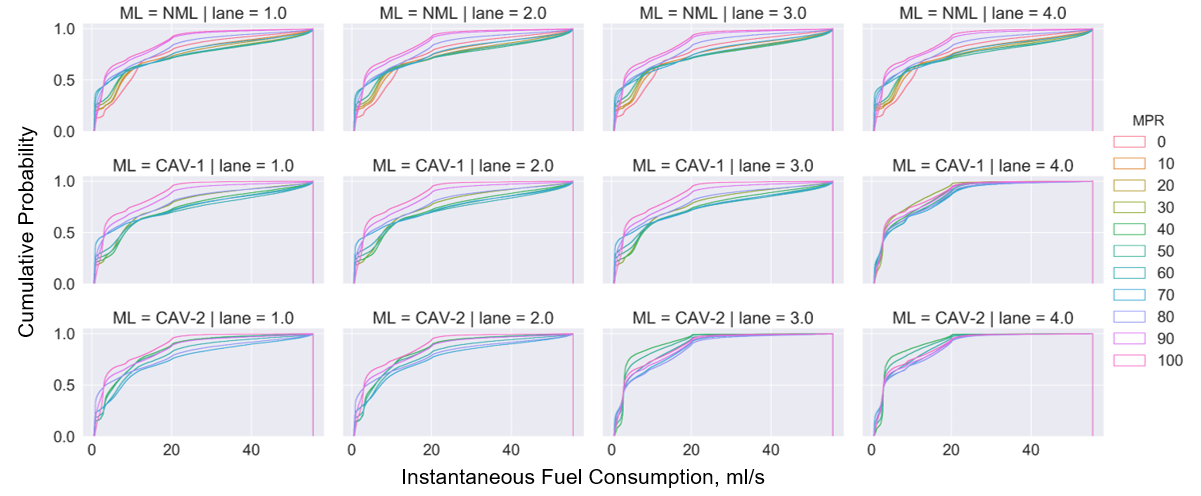}   
	\caption{Instantaneous Fuel Consumption for All Vehicles} 	
	\label{fig:cdfFuelAll}
\end{figure*}

The vehicle data was derived from the raw data from the detectors in three locations (marked in Figure \ref{fig:networkGeometryDemand}). The result for the fuel consumption is plotted in Figure \ref{fig:cdfFuelAll}, which shows two distinctive patterns for the GPLs and the managed lane. The concentration of fuel consumption is within 5 $ml/s$ to 10 $ml/s$ for lanes that allows HV operation (i.e., mixed traffic), when the MPR for CAVs is equal or less than 60\%.  
\textcolor{r1}{
When the MPR rises to above 60\%, the instantaneous fuel consumption shifts to a lower values with a ``narrower'' slope: higher concentration between 5 $ml/s$ and 7 $ml/s$.
}

\textcolor{r1}{
We then isolate the CDF curve for both CAVs and HVs, when they operate on the leftmost lane under homogeneous flow condition. More specifically, the separated CDF curves represent the observations of HVs from the 0\% MPR in NML case and the observations for CAVs from the 100\% MPR for CAV-1 case.  The CDF curves in Figure \ref{fig:cdfFuelAll} exhibit two different patterns for CAVs and HVs. The former with 60\% of the observations fall below 4 $ml/s$, whereas the latter with 60\% of the observations below 12 $ml/s$ with a wider spread. The wider spread for HVs is probably caused by the stochastic nature of human drivers (which is simulated by the Wiedemann model). Hence the mixed traffic condition is comprised of two competing flows that excreting their influence.
}

In the GPLs, the MPR plays as an indicator for the dominance of each traffic flow. The higher the MPR, the closer the CDF curves approach the pattern of managed lane that is used by CAV exclusively. 
In the managed lane, the CAV traffic is the sole dominating traffic. Therefore, the fuel consumption curve exhibits only CAV traffic characteristics, regardless of the MPR. We include the fuel consumption rate CDF curves for HVs and CAVs in Appendix II (Figures \ref{fig:cdfFuelHv} and \ref{fig:cdfFuelCav}) - both figures demonstrate the shift towards CAV fuel consumption CDF pattern as the MPR grows.


%

\begin{figure}[h]
\begin{minipage}[h]{0.5\columnwidth}
\centering
\subfloat[vehicle density]{\label{main:a}\includegraphics[scale=.4]{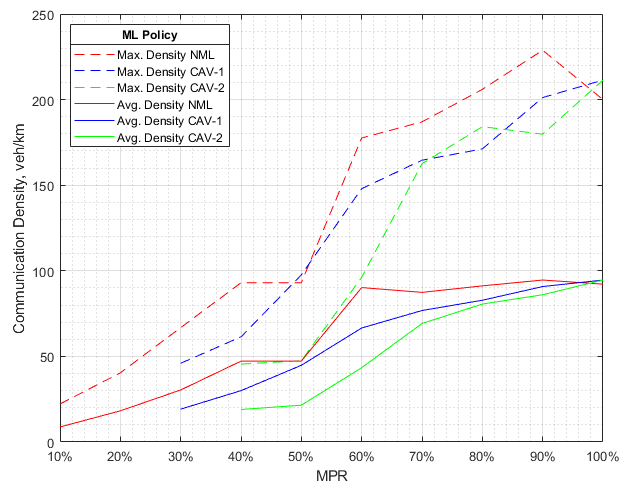}}
\end{minipage}
\begin{minipage}[h]{0.5\columnwidth}
\centering
\subfloat[packet reception rate]{\label{main:b}\includegraphics[scale=.4]{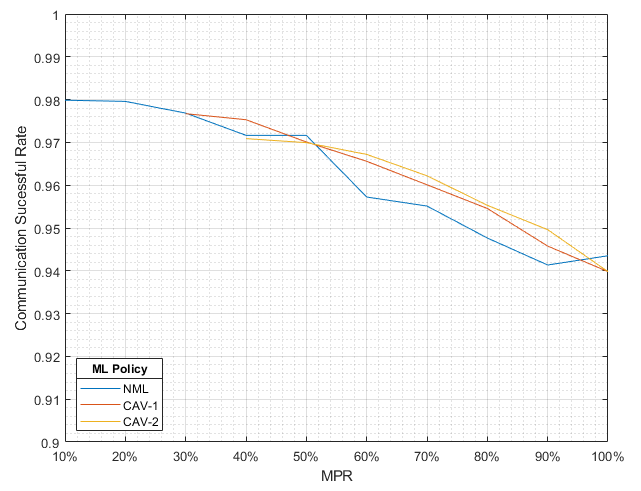}}
\end{minipage}
\caption{V2V communication performance measure}
\label{fig:comm}
\end{figure}

\subsection{Wireless Communication}
Figure \ref{fig:comm}(a) shows the maximum and the average density for instances of V2V communication among three managed lane policies. Recall the DSRC communication model only deals with the physical layer. While the transmission density increases as the MPR increases, the maximum density in NML is higher than CAV-1 and CAV-2, because the CAV platoons were broken down by certain HVs which are susceptible to shockwaves. As such, the traffic flow is compressed, producing a higher traffic density and thus higher transmission density. With the aid of CAV lane, the communication density is thus maintained at a lower level. In a CAV lane, the CAVs distribute longitudinally on the managed lane. The NML, in comparison with two managed lane cases, is more likely to generated pockets of traffic with CAVs across multiple lanes, which could result in localized higher transmission activity.

The probability of successful reception of BSM from a leading vehicle to a subject vehicle is shown in Figure \ref{fig:comm}(b). The probability curves under CAV-1 and CAV-2 scenarios are in close proximity to each other and they are showing the same trend. The maximum difference between these two curves is 0.04 at 90\% MPR.  The probability of successful communication of NML at high MPR range (60\% to 90\%) is consistently lower than those of CAV-1 and CAV-2. This is caused by the compression of traffic flow by localized shockwaves. There is an overall decreasing trend of the probability as the MPR increases, but still remains a successful rate of 94\% and above.

\subsection{Network Performance}
\begin{figure}[h]
	\centering
	\includegraphics[width=0.7\columnwidth]{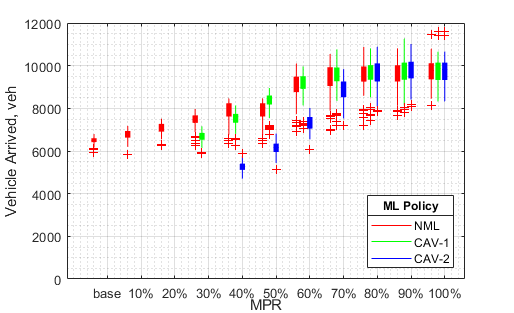}   
	\caption{Network throughput} 
	\label{fig: networkTP}
\end{figure}

The measures used in this section gauges the overall performance of the simulation network at an aggregated level.
The throughput represents the total number of vehicles that have arrived at their destinations, shown in Figure \ref{fig: networkTP}. As mentioned before, the network was configured with a higher than usual demand. With a 10,000-vph demand for a four-lane highway, the network was only able to process 6,500 vph in the absence of CAVs. Under the NML scenario, as the MPR of CAVs increases, so does the network throughput. The throughput reaches approximately 8,000 vph with 40\% and 50\% MPRs. However, at 60\% MPR, the network throughput is boosted again and maintains at the same level at 9,600 vph when the MPR is above 70\%. The throughput in CAV-1 case begins to outperform the NML case at MPR 50\% and keeps increasing to 9700 vph at 70\% MPR, where the throughput starts leveling in spite of the increase in MPR. For the CAV-2 policy, the system throughput only reaches the same level of the two counterparts at 70\% MPR due to under-utilization of CAV lanes with low MRPs.  

\begin{figure}[h]
\begin{minipage}[h]{.5\columnwidth}
\centering
\subfloat[average delay]{\includegraphics[scale=.55]{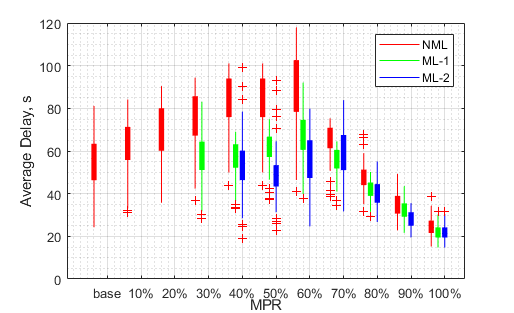}}
\end{minipage} 
\begin{minipage}[h]{.5\columnwidth}
\centering
\subfloat[average speed]{\includegraphics[scale=.55]{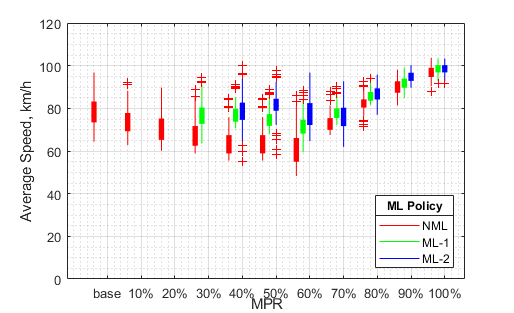}}
\end{minipage} 
\caption{Average Speed and Delay}
\label{fig:delaySpd}
\end{figure}

The average delay experienced by vehicles (plotted in Figure \ref{fig:delaySpd}(a)) within the network is calculated by dividing the total delay by the sum of the vehicles within the network and the vehicles that have exited the network. For three strategies, the average delay starts to decrease as the throughput levels off: at 60\% for NML and CAV-1, and at 70 \% for CAV-2. 
Such seemingly counter-intuitive phenomena could be explained by taking into account the average speed, which is shown in Figure \ref{fig:delaySpd}(b): when the throughput is in a graduate increase as the MPR goes until 60\%, the average speed exhibits a decreasing trend, which is in an inverse relationship with vehicle delay. This trend is in agreement with the speed-flow fundamental diagram.

\section{Discussion and Conclusions}
\label{sect:conclusion}

In this section, we highlight the findings from the previous section and discuss the study in a boarder context.
\subsection{Summary}

The analysis results indicate that the introduction of CAV could increase the throughput of the overall system, even when no managed lane policy is in place. \textcolor{r2}{The congestion region in the speed-flow diagram disappears as the MPR of the CAVs increases. This is an indication of the improvement of roadway capacity owing to CAVs, which is consistent with the findings of previous studies. More importantly, the congestion region first disappears in the CAV lane in CAV-1 case, illustrating that the homogeneity of CAV traffic results in a more stable traffic flow with a high throughput. A CAV lane, with an MPR of as low as 40\%, is able to accommodate more traffic compared to a GP lane and it helps to alleviate the overall congestion of the network.}
The average vehicle delay exhibits a decreasing trend, even after the network throughput levels after 70\% MPR. This is an indicator that the network is able to carry additional traffic than the high demand specified in Fig. \ref{fig:networkGeometryDemand}.

\textcolor{r2}{The individual headways among consecutive vehicles are measured for each lane. From the headway distribution, one can not only measure the compactness of the traffic but also the stability of the traffic flow.
Both HVs and CAVs have a predominate headway as shown in Table \ref{table:vehControl}.  
}
In a heterogeneous traffic flow, two spikes with different tipping points can be observed in the headway CDF curve. Each segments indicates a high concentration for the headway samples. One is for the following headway samples observed on HVs, and the other is for the headway samples for CAVs. With traffic homogeneity on the CAV lane, there is only one spike on the CDF curves. The magnitude of the spike also depends on the lane occupancy, as evidenced by the comparison of CAV-1 and CAV-2 at the same MPR. The two-spike pattern remains even at high-range MPR (i.e., 60-80\%) in the absence of CAV lane (the NML case). 

\textcolor{r2}{The VT-Micro model, which produces instantaneous fuel consumption for individual vehicles, was employed to estimate the environmental impact of the CAV lane. The vehicle speed and acceleration was collected as inputs and the relative fuel consumption, instead of the absolute one, is examined.  Again, distinct patterns for a GPL and a CAV lane were observed. The average instantaneous fuel consumption for CAV lane has a narrower distribution.}

\textcolor{r2}{
Lastly, the DSRC communication was measured using an analytical communication model that is derived from a package-level network simulator. It simulates the physical layer of the DSRC communication that is an integral element of CAVs. We found a lower communication density in CAV lane, as the CAVs were more evenly distributed longitudinally. A lower communication density indicates a less congested communication channel, which increases the performance of the V2V communication.} Compared to CAV-1 and CAV-2 scenarios, it is more likely under NML scenario to generate pockets of traffic with CAVs across multiple lanes, which could introduce higher localized transmission activity and increase the loss of BSM packets.

\textcolor{r2}{The overall results show that a single CAV lane in a four-lane highway network is able to provide the necessary technical accommodation efficiently in the mixed traffic conditions with a wide range of MPR. A CAV dedicated lane is helpful to guarantee the benefits of CAVs, as it creates a homogeneous CAV flow. Implementing two CAV lanes, however, may adversely affect the overall traffic, especially when the MPR of CAV does not warrant an additional CAV lane. }

\subsection{Limitations}
While the paper demonstrates the benefits of managed lane for CAV at lane- and vehicle- levels, we should note that there are limitations in this study and the benefits are realized in a controlled environment under certain assumptions. 
\textcolor{r2}{
First of all, although the Wiedemann model is behaviorally sound and has been adopted by numerous researchers for simulating human drivers, the complexity of a human drivers under dynamic traffic conditions is difficult, if possible at all, to be captured by simulation models. In addition, the behavioral adaption for human drivers in the presence of CAV is not known yet, due to the lack of empirical evidence in the public domain. Preliminary results revealed that a smaller time headway was adopted by a HV when driving along side with closed platooned CAVs \cite{GOUY2014264}. Note that the Wiedemann driver behavior parameters were calibrated using field data where CAVs have not been deployed on the roadway yet. The calibrated parameters represents a subset of the driving population, and they may not directly transferable to other driving conditions or demographics.
The E-IDM, while being widely adopted, does not contain the multi-anticipative car-following feature, which has been promoted as one of the crucial feature enabled by V2V communication. Therefore, the performance of the CAVs are expected to be more conservative. Like many existing CAV car-following models, the E-IDM does not factor in the aspects of human factor that is anticipated to be more pronounced in the lower levels of automation.
}

\textcolor{r2}{
In addition, there are several salient issues regarding the low level automation and its modeling as well. For a CAV, the drivers' acceptance of short following headway (e.g., 0.6 s) is still an open question \cite{Nowakowski2011}, given that the short following headway is technologically attainable. It is reasonable to expect that the acceptance of extremely short headway would be low initially, although it will gradually increase as CAV penetration increases. The pace of adaption, though, is largely depending of the level of confidence to the ADS from human drivers. 
The level of compliance from drivers (in the absence of automation) is also an important factor in harnessing rich information brought by the connectivity. The layer of driver stocatistcity in reacting to traffic information remains. In the extreme case, a complete disregard of useful information could negate the benefits of connectivity. 
}

\textcolor{r2}{Another crucial issue is the transition of control from the ADS back to the human driver. As per the definition of vehicle automation by the SAE, the Level 3 automation (and below) requires a fallback receptive driver when the ADS exits its designed operational domain. As studies have shown, such fallback process is way more complicated than merely re-taking the steering wheel. First, a driver needs to regain situational awareness of the traffic environment from the disengagement of driving. The surge in cognitive demand during the initial period of re-engaging in driving tasks could result in deterioration in driver's performance (e.g, increased reaction time, inadequate situational awareness). This aspect rarely exists in current CAV models, and much likely it will require an endogenous cognitive models that is able to take into account the driving task demand and the cognitive capacity of human drivers \citep{calvert2020generic}. Therefore, the human-machine interfacing is seldom captured in current simulation model, including the one used in this study. 
}

\subsection{Future Research}

\textcolor{r2}{
The future research would focus on relaxing the assumptions in this study. The first direction is the CAV behavior modeling. Researchers have recently started the incorporation of human factor aspect, such as an extension module in IDM to model driver's responses to advanced traffic information  \citep{sharma2019modelling}, an endogenous model of human cognitive for the transition of control \citep{calvert2020generic}. Such developments offer a great opportunity to introduce human factor in a mixed traffic flow in the future. Second, the inner most lane is generally assigned as the managed lane in current practices, which requires eligible users to merge to access the managed lane, and induces additional demand of lane changing. The access plan (e.g., ingress and egress points of the managed lane, eligibility) requires further study to minimize the negative impacts caused by induced weaving activity. A cost-benefit analysis may also be warranted for comparing managed lane strategies with other emerging technologies, such as vehicle awareness device (VAD), for the near-term deployment of CAV. Some researchers have started the exploration of right-most managed lane in U.S. \citep{li2020high}. Lastly, the characteristics of mixed traffic flow that is anticipated in the near-term deployment of CAV needs further exploration. Especially, the impact of CAVs at individual trajectory level by analyzing high-resolution vehicle trajectory data desires further exploration. 
}

\newpage
\appendix
\setcounter{figure}{0}
\setcounter{table}{0}   

\section{List of Abbreviations}
\begin{table}[h]
\centering
\caption{\textcolor{r1}{List of Abbreviations}} 
\resizebox{0.65\textwidth}{!}
{
\begin{tabular}{p{1 in}|p{3in}} 
\hline  \hline
\textbf{Abbreviation} & \textbf{Definition} \\ \hline 
ADAS & advanced driver-assistance systems \\ \hline
ADS & automated driving systems \\ \hline
ACC & adaptive cruise control \\ \hline
AV & automated vehicles  \\ \hline 
API & application programming interface \\ \hline
BSM & basic safety message \\ \hline
CV & connected vehicles \\ \hline
CAV & connected and automated vehicles\\ \hline
CACC & cooperative adaptive cruise control \\ \hline
CAH &  constant-acceleration heuristic  \\ \hline
CDF &  cumulative probability function \\ \hline
CHEM & comprehensive modal emission model \\ \hline
DSRC & dedicated short-range communication \\ \hline
DLL &  dynamic-link library \\ \hline
DTG &  desired time gap \\ \hline
E-IDM & enhanced intelligent driver model \\ \hline
GPL & general purpose lane \\ \hline
HV & human-driven vehicle  \\  \hline  
HOV & high-occupancy vehicles\\ \hline
IEEE & Institute of Electrical and Electronics Engineers \\ \hline
MPR & market penetration rate\\ \hline
MOVES & motor vehicle emission simulator\\ \hline
PET &  post encroachment time  \\ \hline
SSAM & surrogate safety assessment model\\ \hline
SAE & Society of Automotive Engineers International \\ \hline
SUMO &  simulation of urban mobility  \\ \hline
TTC & time to collision \\ \hline
VAD & vehicle awareness device  \\  \hline
WAVE & wireless ac¬cess in vehicular environment  \\ \hline
\hline  
\end{tabular}
}
\label{table:abbrv}
\end{table}


\section{Coefficients for Wireless Communication Model}
\label{app:coefficent}
\begin{figure}[H]
\begin{minipage}[h]{.45\columnwidth}
\subfloat[Coefficient $h^{(j,k)}_{i}$ in Eq. \ref{eq:dsrcMod}]{
\resizebox{\columnwidth}{!}{%
\begin{tabular}{ccccccc}
\hline
 &  &  & $(j,k)$  &  &  \\
\hline
 &(0,0) & (1,0) & (2,0) & (3,0) & (4,0)  \\ 
\hline
$h^{(j,k)}_{1}$ & 0.0209865 & -9.66304$e$-07  &-1.72786$e$-11  & 5.09506$e$-17  & -7.91921$e$-23 \\
$h^{(j,k)}_{2}$ & 2.24743 & 7.84884$e$-07 & 2.28533$e$-10 & -5.89802$e$-16 & 3.55262$e$-22 \\
$h^{(j,k)}_{3}$ & 2.56426 & 2.82287$e$-05 & -7.09939$e$-10  & 1.34371$e$-15 & -3.01956$e$-22\\
$h^{(j,k)}_{4}$ & 2.41146 & -9.32859$e$05 & 6.77403$e$-10 & -9.64188$e$-16 & 3.69652$e$-23\\ 
\hline
 & (3,1) & (2,1) & (2,2) & (1,1) & (1,2)\\
\hline
$h^{(j,k)}_{1}$ & 3.16577$e$-20 & 2.13587$e$-14 & -5.05716$e$-17 & 4.00928$e$-09 & -1.88707$e$-11 \\
$h^{(j,k)}_{2}$ & 4.07120$e$-19 & -2.66510$e$-13 & 8.64273$e$-17 & -7.31274$e$-08 & 2.98549$e$-10 \\
$h^{(j,k)}_{3}$ & -1.85451$e$-18 & 1.02847$e$-12 & 1.80250$e$-16 & 1.56259$e$-07 & -8.50944$e$-10 \\
$h^{(j,k)}_{4}$ & 1.85043$e$-18 & -1.13894$e$-16 & -4.05333$e$-16 & -2.56738$e$-08 & 6.24415$e$-10 \\
\hline
 & (1,3) & (0,1) & (0,2) & (0,3) & (0,4)\\
\hline
$h^{(j,k)}_{1}$ & 3.25406$e$-14 & 0.000418109 & -4.30875$e$-06 & 1.00775$e$-08 & -7.32254$e$-12 \\
$h^{(j,k)}_{2}$ & -3.24982$e$-13 & 0.00498750 & -7.22232$e$-06 & 1.69755$e$-08 & -2.94381$e$-11\\
$h^{(j,k)}_{3}$ & 7.59094$e$-13 & -0.0227008 & 7.50391$e$-05 & -1.81469$e$-07 & 2.02182$e$-10\\
$h^{(j,k)}_{4}$ &-3.57571$e$-13 & 0.0191490 & -6.92678$e$-07 & 1.79917$e$-07 & -2.07263$e$-10\\
\hline
\end{tabular}
\label{table: hCoef}
}
}
\end{minipage} 
\begin{minipage}[h]{.5\columnwidth}
\centering
\subfloat[PDF of successfully reception (300-m power range)]{\includegraphics[scale=.53]{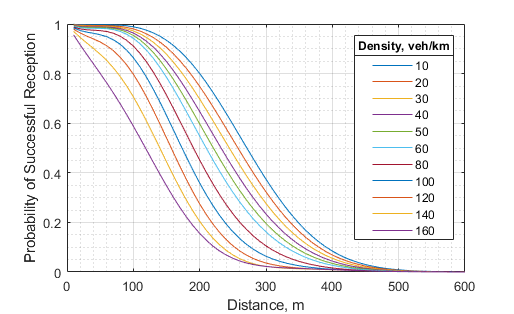}}
\end{minipage} 
\caption{DSRC Model Coefficients \& PDF}
\label{fig:delaySpd}
\end{figure}

The coefficients obtained from the polynomial function $h_{i}(\xi ,\varphi )$ is shown in Table \ref{table: hCoef}. It is worth stressed that even seemingly negligible values, if omitted, could result deviation in the probability of reception from 8\% to 100\% \cite{Killat2009}.

\newpage
\section{Instantaneous Fuel Consumption for HV and CAV}
\label{appdx:fuel}
\begin{figure}[h]
	\centering
	\includegraphics[width=0.8\textwidth]{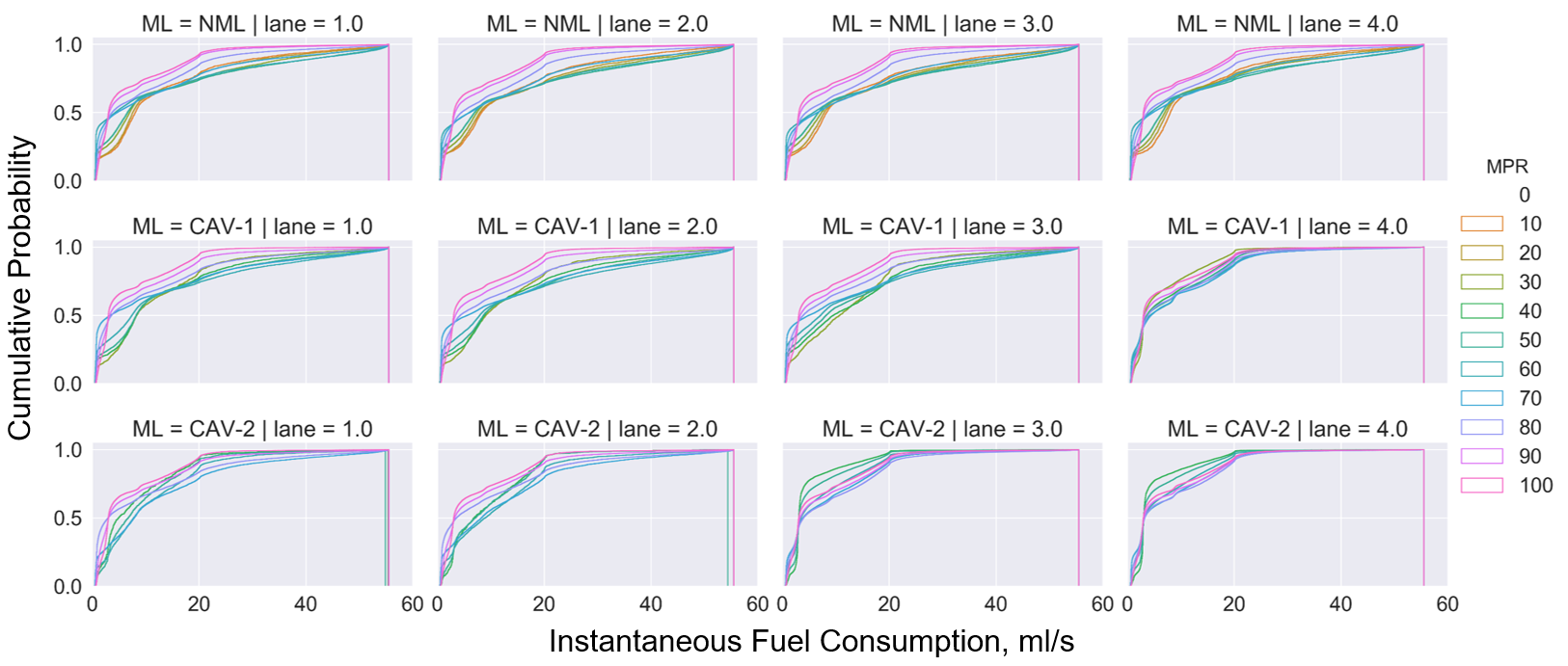}   
	\caption{Instantaneous fuel consumption for HVs} 	
	\label{fig:cdfFuelHv}
\end{figure}

\begin{figure}[h]
	\centering
	\includegraphics[width=0.8\textwidth]{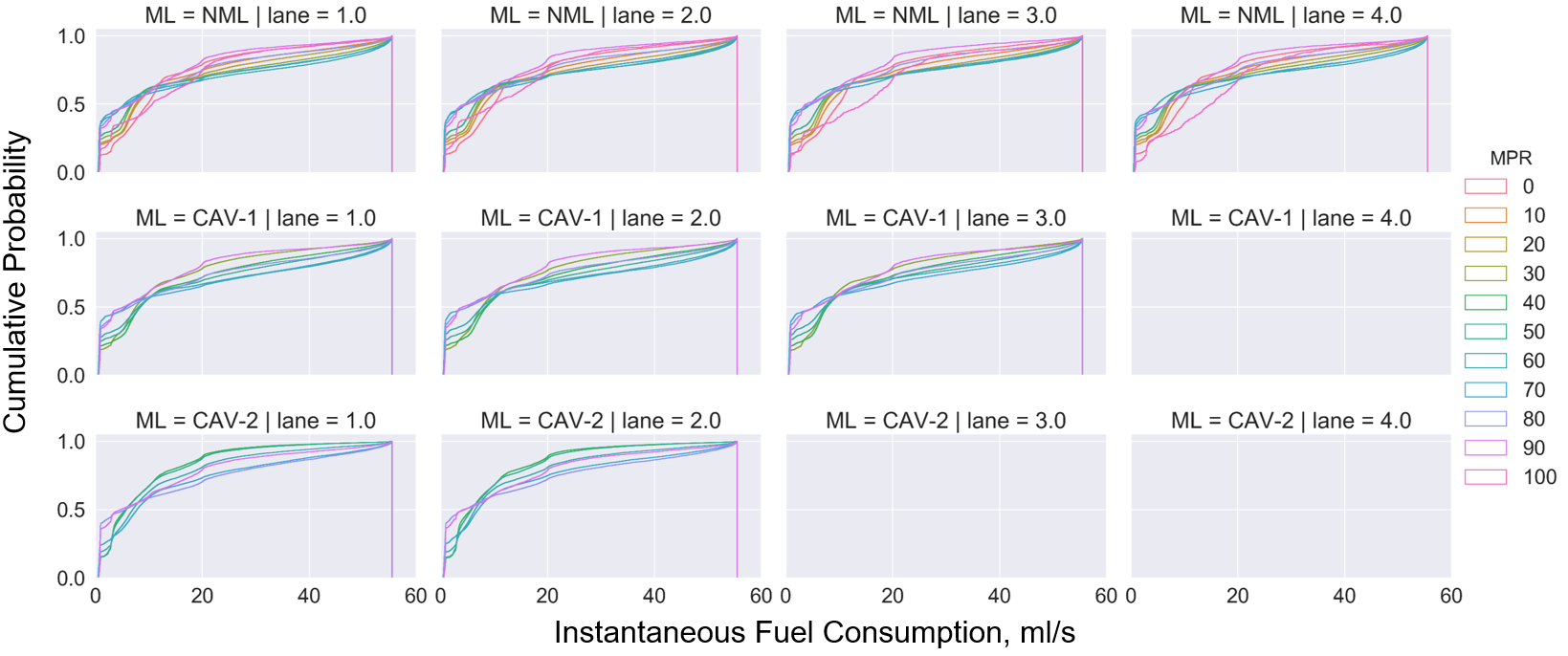}   
	\caption{Instantaneous fuel consumption for CAVs} 	
	\label{fig:cdfFuelCav}
\end{figure}

\begin{figure}[h]
	\centering
	\includegraphics[width=0.5\textwidth]{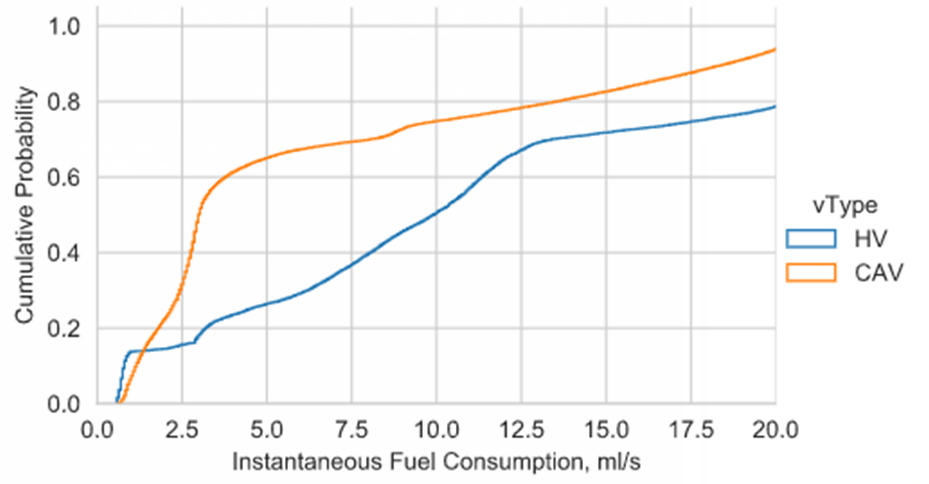}   
	\caption{Instantaneous fuel consumption curve for homogeneous flow} 	
	\label{fig:cdf_homo}
\end{figure}

\section*{Data Availability Statement}
Some or all data, models, or code that support the findings of this study are available from the corresponding author upon reasonable request.

\newpage
\bibliographystyle{elsarticle-num-names}
\bibliography{CAV-ML-Flow_8816540_R2}

\begin{thebibliography}{72}
\expandafter\ifx\csname natexlab\endcsname\relax\def\natexlab#1{#1}\fi
\providecommand{\url}[1]{\texttt{#1}}
\providecommand{\href}[2]{#2}
\providecommand{\path}[1]{#1}
\providecommand{\DOIprefix}{doi:}
\providecommand{\ArXivprefix}{arXiv:}
\providecommand{\URLprefix}{URL: }
\providecommand{\Pubmedprefix}{pmid:}
\providecommand{\doi}[1]{\href{http://dx.doi.org/#1}{\path{#1}}}
\providecommand{\Pubmed}[1]{\href{pmid:#1}{\path{#1}}}
\providecommand{\bibinfo}[2]{#2}
\ifx\xfnm\relax \def\xfnm[#1]{\unskip,\space#1}\fi
\bibitem[{Chan(2017)}]{chan2017advancements}
\bibinfo{author}{C.-Y. Chan},
\newblock \bibinfo{title}{Advancements, prospects, and impacts of automated
  driving systems},
\newblock \bibinfo{journal}{International journal of transportation science and
  technology} \bibinfo{volume}{6} (\bibinfo{year}{2017})
  \bibinfo{pages}{208--216}.
\bibitem[{fhw(2008)}]{fhwa2008Managed}
\bibinfo{title}{{Managed Lanes: A Primer}}, \bibinfo{type}{Technical Report},
  \bibinfo{year}{2008}.
\bibitem[{Kalra and Paddock(2016)}]{Kalra2016}
\bibinfo{author}{N.~Kalra}, \bibinfo{author}{S.~M. Paddock},
\newblock \bibinfo{title}{{Driving to safety: How many miles of driving would
  it take to demonstrate autonomous vehicle reliability?}},
\newblock \bibinfo{journal}{Transportation Research Part A: Policy and
  Practice} \bibinfo{volume}{94} (\bibinfo{year}{2016})
  \bibinfo{pages}{182--193}. \DOIprefix\doi{10.1016/j.tra.2016.09.010}.
\bibitem[{Smith et~al.(2015)Smith, Bellone, Bransfield, Ingles, Noel, Reed,
  Yanagisawa et~al.}]{smith2015benefits}
\bibinfo{author}{S.~Smith}, \bibinfo{author}{J.~Bellone},
  \bibinfo{author}{S.~Bransfield}, \bibinfo{author}{A.~Ingles},
  \bibinfo{author}{G.~Noel}, \bibinfo{author}{E.~Reed},
  \bibinfo{author}{M.~Yanagisawa}, et~al., \bibinfo{title}{Benefits estimation
  framework for automated vehicle operations.}, \bibinfo{type}{Technical
  Report}, United States. Department of Transportation. Intelligent
  Transportation~…, \bibinfo{year}{2015}.
\bibitem[{Hussain et~al.(2016)Hussain, Ghiasi, and Li}]{hussain2016freeway}
\bibinfo{author}{O.~Hussain}, \bibinfo{author}{A.~Ghiasi},
  \bibinfo{author}{X.~Li},
\newblock \bibinfo{title}{Freeway lane management approach in mixed traffic
  environment with connected autonomous vehicles},
\newblock \bibinfo{journal}{arXiv preprint arXiv:1609.02946}
  (\bibinfo{year}{2016}).
\bibitem[{Wang et~al.(2017)Wang, Li, and Work}]{wang2017Comparing}
\bibinfo{author}{R.~Wang}, \bibinfo{author}{Y.~Li}, \bibinfo{author}{D.~B.
  Work},
\newblock \bibinfo{title}{Comparing traffic state estimators for mixed human
  and automated traffic flows},
\newblock \bibinfo{journal}{Transportation Research Part C: Emerging
  Technologies} \bibinfo{volume}{78} (\bibinfo{year}{2017}) \bibinfo{pages}{95
  -- 110}. \URLprefix
  \url{http://www.sciencedirect.com/science/article/pii/S0968090X17300517}.
  \DOIprefix\doi{https://doi.org/10.1016/j.trc.2017.02.011}.
\bibitem[{Van~Arem et~al.(2006)Van~Arem, Van~Driel, and Visser}]{van2006impact}
\bibinfo{author}{B.~Van~Arem}, \bibinfo{author}{C.~J. Van~Driel},
  \bibinfo{author}{R.~Visser},
\newblock \bibinfo{title}{The impact of cooperative adaptive cruise control on
  traffic-flow characteristics},
\newblock \bibinfo{journal}{IEEE Transactions on intelligent transportation
  systems} \bibinfo{volume}{7} (\bibinfo{year}{2006})
  \bibinfo{pages}{429--436}.
\bibitem[{Shladover et~al.(2012)Shladover, Su, and Lu}]{shladover2012impacts}
\bibinfo{author}{S.~E. Shladover}, \bibinfo{author}{D.~Su},
  \bibinfo{author}{X.-Y. Lu},
\newblock \bibinfo{title}{Impacts of cooperative adaptive cruise control on
  freeway traffic flow},
\newblock \bibinfo{journal}{Transportation Research Record}
  \bibinfo{volume}{2324} (\bibinfo{year}{2012}) \bibinfo{pages}{63--70}.
\bibitem[{{Hall} and {Tsao}(1997)}]{Hall1997capaciy}
\bibinfo{author}{R.~W. {Hall}}, \bibinfo{author}{H.~.~J. {Tsao}},
\newblock \bibinfo{title}{Capacity of automated highway systems: merging
  efficiency},
\newblock in: \bibinfo{booktitle}{Proceedings of the 1997 American Control
  Conference (Cat. No.97CH36041)}, volume~\bibinfo{volume}{3},
  \bibinfo{year}{1997}, pp. \bibinfo{pages}{2046--2050 vol.3}.
  \DOIprefix\doi{10.1109/ACC.1997.611049}.
\bibitem[{Arnaout and Arnaout(2014)}]{Arnaout2014}
\bibinfo{author}{G.~Arnaout}, \bibinfo{author}{J.-P. Arnaout},
\newblock \bibinfo{title}{{Exploring the effects of cooperative adaptive cruise
  control on highway traffic flow using microscopic traffic simulation}},
\newblock \bibinfo{journal}{Transportation Planning and Technology}
  (\bibinfo{year}{2014}). \DOIprefix\doi{10.1080/03081060.2013.870791}.
\bibitem[{Songchitruksa et~al.(2016)Songchitruksa, Bibeka, Lin, Zhang
  et~al.}]{songchitruksa2016}
\bibinfo{author}{P.~Songchitruksa}, \bibinfo{author}{A.~Bibeka},
  \bibinfo{author}{L.~I. Lin}, \bibinfo{author}{Y.~Zhang}, et~al.,
  \bibinfo{title}{Incorporating driver behaviors into connected and automated
  vehicle simulation.}, \bibinfo{type}{Technical Report}, Center for Advancing
  Transportation Leadership and Safety (ATLAS Center), \bibinfo{year}{2016}.
\bibitem[{Lee et~al.(2014)Lee, Bared, and Park}]{lee2014mobility}
\bibinfo{author}{J.~Lee}, \bibinfo{author}{J.~Bared},
  \bibinfo{author}{B.~Park},
\newblock \bibinfo{title}{Mobility impacts of cooperative adaptive cruise
  control (cacc) under mixed traffic conditions},
\newblock in: \bibinfo{booktitle}{93rd Annual Meeting of the Transportation
  Research Board, Washington, DC}, \bibinfo{year}{2014}.
\bibitem[{Liu et~al.(2020)Liu, Shladover, Lu, and Kan}]{Liu2019Freeway}
\bibinfo{author}{H.~Liu}, \bibinfo{author}{S.~E. Shladover},
  \bibinfo{author}{X.-Y. Lu}, \bibinfo{author}{X.~D. Kan},
\newblock \bibinfo{title}{Freeway vehicle fuel efficiency improvement via
  cooperative adaptive cruise control},
\newblock \bibinfo{journal}{Journal of Intelligent Transportation Systems}
  \bibinfo{volume}{0} (\bibinfo{year}{2020}) \bibinfo{pages}{1--13}.
  \DOIprefix\doi{10.1080/15472450.2020.1720673}.
\bibitem[{Liu et~al.(2018)Liu, Kan, Shladover, Lu, and
  Ferlis}]{Liu2018Modeling}
\bibinfo{author}{H.~Liu}, \bibinfo{author}{X.~D. Kan}, \bibinfo{author}{S.~E.
  Shladover}, \bibinfo{author}{X.-Y. Lu}, \bibinfo{author}{R.~E. Ferlis},
\newblock \bibinfo{title}{Modeling impacts of cooperative adaptive cruise
  control on mixed traffic flow in multi-lane freeway facilities},
\newblock \bibinfo{journal}{Transportation Research Part C: Emerging
  Technologies} \bibinfo{volume}{95} (\bibinfo{year}{2018}) \bibinfo{pages}{261
  -- 279}. \URLprefix
  \url{http://www.sciencedirect.com/science/article/pii/S0968090X18310313}.
  \DOIprefix\doi{https://doi.org/10.1016/j.trc.2018.07.027}.
\bibitem[{An et~al.(1997)An, Barth, Norbeck, and Ross}]{an1997development}
\bibinfo{author}{F.~An}, \bibinfo{author}{M.~Barth},
  \bibinfo{author}{J.~Norbeck}, \bibinfo{author}{M.~Ross},
\newblock \bibinfo{title}{Development of comprehensive modal emissions model:
  operating under hot-stabilized conditions},
\newblock \bibinfo{journal}{Transportation Research Record}
  \bibinfo{volume}{1587} (\bibinfo{year}{1997}) \bibinfo{pages}{52--62}.
\bibitem[{Rakha et~al.(2004)Rakha, Ahn, and Trani}]{Rakha2004}
\bibinfo{author}{H.~Rakha}, \bibinfo{author}{K.~Ahn},
  \bibinfo{author}{A.~Trani},
\newblock \bibinfo{title}{{Development of VT-Micro model for estimating hot
  stabilized light duty vehicle and truck emissions}},
\newblock \bibinfo{journal}{Transportation Research Part D: Transport and
  Environment} \bibinfo{volume}{9} (\bibinfo{year}{2004})
  \bibinfo{pages}{49--74}. \URLprefix
  \url{http://www.sciencedirect.com/science/article/pii/S1361920903000543}.
  \DOIprefix\doi{http://dx.doi.org/10.1016/S1361-9209(03)00054-3}.
\bibitem[{Brooker et~al.(2015)Brooker, Gonder, Wang, Wood, Lopp, and
  Ramroth}]{brooker2015fastsim}
\bibinfo{author}{A.~Brooker}, \bibinfo{author}{J.~Gonder},
  \bibinfo{author}{L.~Wang}, \bibinfo{author}{E.~Wood},
  \bibinfo{author}{S.~Lopp}, \bibinfo{author}{L.~Ramroth},
  \bibinfo{title}{FASTSim: A model to estimate vehicle efficiency, cost and
  performance}, \bibinfo{type}{Technical Report}, SAE Technical Paper,
  \bibinfo{year}{2015}.
\bibitem[{Casey and Lund(1992)}]{casey1992changes}
\bibinfo{author}{S.~M. Casey}, \bibinfo{author}{A.~K. Lund},
\newblock \bibinfo{title}{Changes in speed and speed adaptation following
  increase in national maximum speed limit},
\newblock \bibinfo{journal}{Journal of Safety Research} \bibinfo{volume}{23}
  (\bibinfo{year}{1992}) \bibinfo{pages}{135 -- 146}. \URLprefix
  \url{http://www.sciencedirect.com/science/article/pii/0022437592900163}.
  \DOIprefix\doi{https://doi.org/10.1016/0022-4375(92)90016-3}.
\bibitem[{Nowakowski et~al.(2011)Nowakowski, Shladover, Cody, Bu, O'Connell,
  Spring, Dickey, and Nelson}]{nowakowski2011cooperative}
\bibinfo{author}{C.~Nowakowski}, \bibinfo{author}{S.~E. Shladover},
  \bibinfo{author}{D.~Cody}, \bibinfo{author}{F.~Bu},
  \bibinfo{author}{J.~O'Connell}, \bibinfo{author}{J.~Spring},
  \bibinfo{author}{S.~Dickey}, \bibinfo{author}{D.~Nelson},
  \bibinfo{title}{Cooperative adaptive cruise control: Testing drivers' choices
  of following distances}, \bibinfo{type}{Technical Report},
  \bibinfo{year}{2011}.
\bibitem[{Gouy et~al.(2014)Gouy, Wiedemann, Stevens, Brunett, and
  Reed}]{GOUY2014264}
\bibinfo{author}{M.~Gouy}, \bibinfo{author}{K.~Wiedemann},
  \bibinfo{author}{A.~Stevens}, \bibinfo{author}{G.~Brunett},
  \bibinfo{author}{N.~Reed},
\newblock \bibinfo{title}{Driving next to automated vehicle platoons: How do
  short time headways influence non-platoon drivers’ longitudinal control?},
\newblock \bibinfo{journal}{Transportation Research Part F: Traffic Psychology
  and Behaviour} \bibinfo{volume}{27} (\bibinfo{year}{2014})
  \bibinfo{pages}{264 -- 273}. \URLprefix
  \url{http://www.sciencedirect.com/science/article/pii/S1369847814000345}.
  \DOIprefix\doi{https://doi.org/10.1016/j.trf.2014.03.003}.
\bibitem[{Fairclough et~al.(1997)Fairclough, May, and
  Carter}]{FAIRCLOUGH1997387}
\bibinfo{author}{S.~H. Fairclough}, \bibinfo{author}{A.~J. May},
  \bibinfo{author}{C.~Carter},
\newblock \bibinfo{title}{The effect of time headway feedback on following
  behaviour},
\newblock \bibinfo{journal}{Accident Analysis \& Prevention}
  \bibinfo{volume}{29} (\bibinfo{year}{1997}) \bibinfo{pages}{387 -- 397}.
  \URLprefix
  \url{http://www.sciencedirect.com/science/article/pii/S0001457597000055}.
  \DOIprefix\doi{https://doi.org/10.1016/S0001-4575(97)00005-5}.
\bibitem[{Shewmake and Jarvis(2014)}]{shewmake2014hybrid}
\bibinfo{author}{S.~Shewmake}, \bibinfo{author}{L.~Jarvis},
\newblock \bibinfo{title}{Hybrid cars and hov lanes},
\newblock \bibinfo{journal}{Transportation Research Part A: Policy and
  Practice} \bibinfo{volume}{67} (\bibinfo{year}{2014})
  \bibinfo{pages}{304--319}.
\bibitem[{Chang et~al.(2008)Chang, Wiegmann, Smith, Bilotto
  et~al.}]{chang2008review}
\bibinfo{author}{M.~Chang}, \bibinfo{author}{J.~Wiegmann},
  \bibinfo{author}{A.~Smith}, \bibinfo{author}{C.~Bilotto}, et~al.,
  \bibinfo{title}{A review of HOV lane performance and policy options in the
  United States}, \bibinfo{type}{Technical Report}, United States. Federal
  Highway Administration, \bibinfo{year}{2008}.
\bibitem[{Gomez-Ibanez et~al.(2018)Gomez-Ibanez, Casady, Fagan, Foote, and
  Marsh}]{gomez2018toll}
\bibinfo{author}{J.~A. Gomez-Ibanez}, \bibinfo{author}{C.~B. Casady},
  \bibinfo{author}{M.~Fagan}, \bibinfo{author}{J.~Foote},
  \bibinfo{author}{E.~Marsh}, \bibinfo{title}{Toll-Managed Lanes: Benefit-Cost
  Analyses of Seven Projects}, \bibinfo{type}{Technical Report},
  \bibinfo{year}{2018}.
\bibitem[{Nowakowski et~al.(2011)Nowakowski, Shladover, and
  Cody}]{Nowakowski2011}
\bibinfo{author}{C.~Nowakowski}, \bibinfo{author}{S.~E. Shladover},
  \bibinfo{author}{D.~Cody},
\newblock \bibinfo{title}{{Cooperative Adaptive Cruise Control : Testing
  Drivers' Choices of Following Distances}}  (\bibinfo{year}{2011}).
  \DOIprefix\doi{UCB-ITS-PRR-2010-39}.
\bibitem[{Shladover et~al.(2018)Shladover, Nowakowski, and
  Lu}]{shladover2018using}
\bibinfo{author}{S.~E. Shladover}, \bibinfo{author}{C.~Nowakowski},
  \bibinfo{author}{X.-Y. Lu}, \bibinfo{title}{{Using Cooperative Adaptive
  Cruise Control ( CACC ) to Form High-Performance Vehicle Streams Definitions
  , Literature Review and Operational Concept Alternatives}},
  \bibinfo{type}{Technical Report}, University of California Berkeley:
  California Partners for Advanced Transportation Technology,
  \bibinfo{year}{2018}. \URLprefix
  \url{https://escholarship.org/uc/item/3w6920wz}.
\bibitem[{Swaroop et~al.(1994)Swaroop, Hedrick, Chien, and
  Ioannou}]{swaroop1994comparision}
\bibinfo{author}{D.~Swaroop}, \bibinfo{author}{J.~K. Hedrick},
  \bibinfo{author}{C.~Chien}, \bibinfo{author}{P.~Ioannou},
\newblock \bibinfo{title}{A comparision of spacing and headway control laws for
  automatically controlled vehicles1},
\newblock \bibinfo{journal}{Vehicle System Dynamics} \bibinfo{volume}{23}
  (\bibinfo{year}{1994}) \bibinfo{pages}{597--625}.
\bibitem[{Wang et~al.(2019)Wang, Qin, Wang, and Chen}]{Wang2019stability}
\bibinfo{author}{H.~Wang}, \bibinfo{author}{Y.~Qin}, \bibinfo{author}{W.~Wang},
  \bibinfo{author}{J.~Chen},
\newblock \bibinfo{title}{Stability of {CACC}-manual heterogeneous vehicular
  flow with partial {CACC} performance degrading},
\newblock \bibinfo{journal}{Transportmetrica B: Transport Dynamics}
  \bibinfo{volume}{7} (\bibinfo{year}{2019}) \bibinfo{pages}{788--813}.
  \DOIprefix\doi{10.1080/21680566.2018.1517058}.
\bibitem[{Shladover et~al.(2015)Shladover, Nowakowski, Lu, and
  Ferlis}]{shladover2015cooperative}
\bibinfo{author}{S.~E. Shladover}, \bibinfo{author}{C.~Nowakowski},
  \bibinfo{author}{X.-Y. Lu}, \bibinfo{author}{R.~Ferlis},
\newblock \bibinfo{title}{Cooperative adaptive cruise control: Definitions and
  operating concepts},
\newblock \bibinfo{journal}{Transportation Research Record: Journal of the
  Transportation Research Board}  (\bibinfo{year}{2015})
  \bibinfo{pages}{145--152}.
\bibitem[{Larson et~al.(2013)Larson, Kammer, Liang, and
  Johansson}]{larson2013coordinated}
\bibinfo{author}{J.~Larson}, \bibinfo{author}{C.~Kammer},
  \bibinfo{author}{K.-Y. Liang}, \bibinfo{author}{K.~H. Johansson},
\newblock \bibinfo{title}{Coordinated route optimization for heavy-duty vehicle
  platoons},
\newblock in: \bibinfo{booktitle}{Intelligent Transportation Systems-(ITSC),
  2013 16th International IEEE Conference on}, \bibinfo{organization}{IEEE},
  \bibinfo{year}{2013}, pp. \bibinfo{pages}{1196--1202}.
\bibitem[{Kuutti et~al.(2018)Kuutti, Fallah, Katsaros, Dianati, Mccullough, and
  Mouzakitis}]{kuutti2018survey}
\bibinfo{author}{S.~Kuutti}, \bibinfo{author}{S.~Fallah},
  \bibinfo{author}{K.~Katsaros}, \bibinfo{author}{M.~Dianati},
  \bibinfo{author}{F.~Mccullough}, \bibinfo{author}{A.~Mouzakitis},
\newblock \bibinfo{title}{A survey of the state-of-the-art localization
  techniques and their potentials for autonomous vehicle applications},
\newblock \bibinfo{journal}{IEEE Internet of Things Journal}
  \bibinfo{volume}{5} (\bibinfo{year}{2018}) \bibinfo{pages}{829--846}.
\bibitem[{Zhong(2018)}]{ZhongDis}
\bibinfo{author}{Z.~Zhong}, \bibinfo{title}{Assessing the Effectiveness of
  Managed Lane Strategies for the Rapid Deployment of Cooperative Adaptive
  Cruise Control Technology}, Ph.D. thesis, \bibinfo{address}{Newark, NJ},
  \bibinfo{year}{2018}. \URLprefix
  \url{http://archives.njit.edu/vol01/etd/2010s/2018/njit-etd2018-033/njit-etd2018-033.pdf}.
\bibitem[{Zhang et~al.(2018)Zhang, Ma, Smith, and Liu}]{zhang2018operational}
\bibinfo{author}{X.~Zhang}, \bibinfo{author}{J.~Ma},
  \bibinfo{author}{B.~Smith}, \bibinfo{author}{J.~Liu},
  \bibinfo{title}{Operational Performance Evaluation of the Managed Lane
  Strategy for Early Deployment of Cooperative Adaptive Cruise Control},
  \bibinfo{type}{Technical Report}, \bibinfo{year}{2018}.
\bibitem[{Wright et~al.(2018)Wright, Horowitz, and
  Kurzhanskiy}]{wright2018dynamic}
\bibinfo{author}{M.~A. Wright}, \bibinfo{author}{R.~Horowitz},
  \bibinfo{author}{A.~A. Kurzhanskiy},
\newblock \bibinfo{title}{A dynamic-system-based approach to modeling driver
  movements across general-purpose/managed lane interfaces},
\newblock in: \bibinfo{booktitle}{ASME 2018 Dynamic Systems and Control
  Conference}, \bibinfo{organization}{American Society of Mechanical
  Engineers}, \bibinfo{year}{2018}, pp.
  \bibinfo{pages}{V002T15A003--V002T15A003}.
\bibitem[{Chen et~al.(2016)Chen, He, Zhang, and Yin}]{chen2016optimal}
\bibinfo{author}{Z.~Chen}, \bibinfo{author}{F.~He}, \bibinfo{author}{L.~Zhang},
  \bibinfo{author}{Y.~Yin},
\newblock \bibinfo{title}{Optimal deployment of autonomous vehicle lanes with
  endogenous market penetration},
\newblock \bibinfo{journal}{Transportation Research Part C: Emerging
  Technologies} \bibinfo{volume}{72} (\bibinfo{year}{2016})
  \bibinfo{pages}{143--156}.
\bibitem[{Zhong and Lee(2019)}]{Zhong2019Effectiveness}
\bibinfo{author}{Z.~Zhong}, \bibinfo{author}{J.~Lee},
\newblock \bibinfo{title}{{The effectiveness of managed lane strategies for the
  near-term deployment of cooperative adaptive cruise control}},
\newblock \bibinfo{journal}{Transportation Research Part A: Policy and
  Practice} \bibinfo{volume}{129} (\bibinfo{year}{2019})
  \bibinfo{pages}{257--270}. \URLprefix
  \url{http://www.sciencedirect.com/science/article/pii/S0965856419303520}.
  \DOIprefix\doi{https://doi.org/10.1016/j.tra.2019.08.015}.
\bibitem[{Zhong(2018)}]{Zhong2018Assess}
\bibinfo{author}{Z.~Zhong}, \bibinfo{title}{{Assessing the Effectiveness of
  Managed Lane Strategies for the Rapid Deployment of Cooperative Adaptive
  Cruise Control Technology}}, Ph.D. thesis, \bibinfo{year}{2018}.
\bibitem[{Zhong and Lee(2018)}]{Zhong2018Simulation}
\bibinfo{author}{Z.~Zhong}, \bibinfo{author}{J.~Lee},
\newblock \bibinfo{title}{{Simulation Framework for Cooperative Adaptive Cruise
  Control with Empirical DSRC Module}},
\newblock in: \bibinfo{booktitle}{The 44th Annual Conference of the IEEE
  Industrial Electronics Society}, \bibinfo{address}{Washington DC, United
  States}, \bibinfo{year}{2018}. \URLprefix
  \url{http://arxiv.org/abs/1810.06510}.
  \href{http://arxiv.org/abs/1810.06510}{{\tt arXiv:1810.06510}}.
\bibitem[{Qom et~al.(2016)Qom, Xiao, and Hadi}]{qom2016evaluation}
\bibinfo{author}{S.~F. Qom}, \bibinfo{author}{Y.~Xiao},
  \bibinfo{author}{M.~Hadi},
\newblock \bibinfo{title}{Evaluation of cooperative adaptive cruise control
  (cacc) vehicles on managed lanes utilizing macroscopic and mesoscopic
  simulation},
\newblock in: \bibinfo{booktitle}{Transportation Research Board 95th Annual
  Meeting}, \bibinfo{number}{16-6384}, \bibinfo{year}{2016}.
\bibitem[{Ghiasi et~al.(2017)Ghiasi, Hussain, Qian, and Li}]{ghiasi2017mixed}
\bibinfo{author}{A.~Ghiasi}, \bibinfo{author}{O.~Hussain},
  \bibinfo{author}{Z.~S. Qian}, \bibinfo{author}{X.~Li},
\newblock \bibinfo{title}{A mixed traffic capacity analysis and lane management
  model for connected automated vehicles: A markov chain method},
\newblock \bibinfo{journal}{Transportation Research Part B: Methodological}
  \bibinfo{volume}{106} (\bibinfo{year}{2017}) \bibinfo{pages}{266--292}.
\bibitem[{Zhong et~al.(2017)Zhong, Joyoung, and Zhao}]{zhong2017evaluations}
\bibinfo{author}{Z.~Zhong}, \bibinfo{author}{L.~Joyoung},
  \bibinfo{author}{L.~Zhao},
\newblock \bibinfo{title}{Evaluations of managed lane strategies for arterial
  deployment of cooperative adaptive cruise control},
\newblock in: \bibinfo{booktitle}{96th Transportation Research Board Annual
  Meeting}, \bibinfo{year}{2017}.
\bibitem[{Papadoulis et~al.(2019)Papadoulis, Quddus, and
  Imprialou}]{papadoulis2019evaluating}
\bibinfo{author}{A.~Papadoulis}, \bibinfo{author}{M.~Quddus},
  \bibinfo{author}{M.~Imprialou},
\newblock \bibinfo{title}{Evaluating the safety impact of connected and
  autonomous vehicles on motorways},
\newblock \bibinfo{journal}{Accident Analysis \& Prevention}
  \bibinfo{volume}{124} (\bibinfo{year}{2019}) \bibinfo{pages}{12--22}.
\bibitem[{Zhong et~al.(2020)Zhong, Lee, Nejad, and Lee}]{Zhong2020Influence}
\bibinfo{author}{Z.~Zhong}, \bibinfo{author}{E.~E. Lee},
  \bibinfo{author}{M.~Nejad}, \bibinfo{author}{J.~Lee},
\newblock \bibinfo{title}{{Influence of CAV clustering strategies on mixed
  traffic flow characteristics: An analysis of vehicle trajectory data}},
\newblock \bibinfo{journal}{Transportation Research Part C: Emerging
  Technologies} \bibinfo{volume}{115} (\bibinfo{year}{2020})
  \bibinfo{pages}{102611}. \URLprefix
  \url{http://www.sciencedirect.com/science/article/pii/S0968090X19307648}.
  \DOIprefix\doi{https://doi.org/10.1016/j.trc.2020.102611}.
\bibitem[{Ali et~al.(2018)Ali, Zheng, and Haque}]{ali2018connectivity}
\bibinfo{author}{Y.~Ali}, \bibinfo{author}{Z.~Zheng}, \bibinfo{author}{M.~M.
  Haque},
\newblock \bibinfo{title}{Connectivity’s impact on mandatory lane-changing
  behaviour: Evidences from a driving simulator study},
\newblock \bibinfo{journal}{Transportation Research Part C: Emerging
  Technologies} \bibinfo{volume}{93} (\bibinfo{year}{2018})
  \bibinfo{pages}{292--309}.
\bibitem[{{PTV AG}(2018)}]{ptv2018}
\bibinfo{author}{{PTV AG}},
\newblock \bibinfo{title}{Ptv vissim},
\newblock \bibinfo{journal}{Retrieved from PTV Group: http://vision-traffic.
  ptvgroup. com/enus/products/ptv-vissim/use-cases/junction-geometry}
  (\bibinfo{year}{2018}).
\bibitem[{Kesting et~al.(2010)Kesting, Treiber, and
  Helbing}]{kesting2010enhanced}
\bibinfo{author}{A.~Kesting}, \bibinfo{author}{M.~Treiber},
  \bibinfo{author}{D.~Helbing},
\newblock \bibinfo{title}{Enhanced intelligent driver model to access the
  impact of driving strategies on traffic capacity},
\newblock \bibinfo{journal}{Philosophical Transactions of the Royal Society A:
  Mathematical, Physical and Engineering Sciences} \bibinfo{volume}{368}
  (\bibinfo{year}{2010}) \bibinfo{pages}{4585--4605}.
\bibitem[{Wang et~al.(2019)Wang, van Maarseveen, Happee, Tool, and van
  Arem}]{Wang2019Benefits}
\bibinfo{author}{M.~Wang}, \bibinfo{author}{S.~van Maarseveen},
  \bibinfo{author}{R.~Happee}, \bibinfo{author}{O.~Tool},
  \bibinfo{author}{B.~van Arem},
\newblock \bibinfo{title}{Benefits and risks of truck platooning on freeway
  operations near entrance ramp},
\newblock \bibinfo{journal}{Transportation Research Record}
  \bibinfo{volume}{2673} (\bibinfo{year}{2019}) \bibinfo{pages}{588--602}.
  \DOIprefix\doi{10.1177/0361198119842821}.
\bibitem[{Kesting et~al.(2008)Kesting, Treiber, Schönhof, and
  Helbing}]{Kesting2008adaptive}
\bibinfo{author}{A.~Kesting}, \bibinfo{author}{M.~Treiber},
  \bibinfo{author}{M.~Schönhof}, \bibinfo{author}{D.~Helbing},
\newblock \bibinfo{title}{Adaptive cruise control design for active congestion
  avoidance},
\newblock \bibinfo{journal}{Transportation Research Part C: Emerging
  Technologies} \bibinfo{volume}{16} (\bibinfo{year}{2008}) \bibinfo{pages}{668
  -- 683}. \URLprefix
  \url{http://www.sciencedirect.com/science/article/pii/S0968090X08000028}.
  \DOIprefix\doi{https://doi.org/10.1016/j.trc.2007.12.004}.
\bibitem[{Talebpour et~al.(2016)Talebpour, Mahmassani, and
  Bustamante}]{Talebpour2016Modeling}
\bibinfo{author}{A.~Talebpour}, \bibinfo{author}{H.~S. Mahmassani},
  \bibinfo{author}{F.~E. Bustamante},
\newblock \bibinfo{title}{Modeling driver behavior in a connected environment:
  Integrated microscopic simulation of traffic and mobile wireless
  telecommunication systems},
\newblock \bibinfo{journal}{Transportation Research Record}
  \bibinfo{volume}{2560} (\bibinfo{year}{2016}) \bibinfo{pages}{75--86}.
  \DOIprefix\doi{10.3141/2560-09}.
\bibitem[{{Spiliopoulou} et~al.(2017){Spiliopoulou}, {Perraki}, {Papageorgiou},
  and {Roncoli}}]{Spiliopoulou2017Exploitation}
\bibinfo{author}{A.~{Spiliopoulou}}, \bibinfo{author}{G.~{Perraki}},
  \bibinfo{author}{M.~{Papageorgiou}}, \bibinfo{author}{C.~{Roncoli}},
\newblock \bibinfo{title}{Exploitation of acc systems towards improved traffic
  flow efficiency on motorways},
\newblock in: \bibinfo{booktitle}{2017 5th IEEE International Conference on
  Models and Technologies for Intelligent Transportation Systems (MT-ITS)},
  \bibinfo{year}{2017}, pp. \bibinfo{pages}{37--43}.
  \DOIprefix\doi{10.1109/MTITS.2017.8005706}.
\bibitem[{Guériau et~al.(2016)Guériau, Billot, Faouzi, Monteil, Armetta, and
  Hassas}]{gueriau2016assess}
\bibinfo{author}{M.~Guériau}, \bibinfo{author}{R.~Billot},
  \bibinfo{author}{N.-E.~E. Faouzi}, \bibinfo{author}{J.~Monteil},
  \bibinfo{author}{F.~Armetta}, \bibinfo{author}{S.~Hassas},
\newblock \bibinfo{title}{How to assess the benefits of connected vehicles? a
  simulation framework for the design of cooperative traffic management
  strategies},
\newblock \bibinfo{journal}{Transportation Research Part C: Emerging
  Technologies} \bibinfo{volume}{67} (\bibinfo{year}{2016}) \bibinfo{pages}{266
  -- 279}. \URLprefix
  \url{http://www.sciencedirect.com/science/article/pii/S0968090X16000462}.
  \DOIprefix\doi{https://doi.org/10.1016/j.trc.2016.01.020}.
\bibitem[{Treiber et~al.(2000)Treiber, Hennecke, and Helbing}]{Treiber2000}
\bibinfo{author}{M.~Treiber}, \bibinfo{author}{A.~Hennecke},
  \bibinfo{author}{D.~Helbing},
\newblock \bibinfo{title}{{Congested Traffic States in Empirical Observations
  and Microscopic Simulations}},
\newblock \bibinfo{journal}{Physical Review E} \bibinfo{volume}{62}
  (\bibinfo{year}{2000}) \bibinfo{pages}{1805}. \URLprefix
  \url{http://link.aps.org/doi/10.1103/PhysRevE.62.1805
  http://www.theo2.physik.uni-stuttgart.de/treiber/}.
  \DOIprefix\doi{10.1103/PhysRevE.62.1805}.
  \href{http://arxiv.org/abs/0002177}{{\tt arXiv:0002177}}.
\bibitem[{Calvert et~al.(2018)Calvert, van Wageningen-Kessels, and
  Hoogendoorn}]{calvert2018capacity}
\bibinfo{author}{S.~C. Calvert}, \bibinfo{author}{F.~L. van
  Wageningen-Kessels}, \bibinfo{author}{S.~P. Hoogendoorn},
\newblock \bibinfo{title}{Capacity drop through reaction times in heterogeneous
  traffic},
\newblock \bibinfo{journal}{Journal of traffic and transportation engineering
  (English edition)} \bibinfo{volume}{5} (\bibinfo{year}{2018})
  \bibinfo{pages}{96--104}.
\bibitem[{Sharma et~al.(2019)Sharma, Zheng, Kim, Bhaskar, and
  Haque}]{sharma2019estimating}
\bibinfo{author}{A.~Sharma}, \bibinfo{author}{Z.~Zheng},
  \bibinfo{author}{J.~Kim}, \bibinfo{author}{A.~Bhaskar},
  \bibinfo{author}{M.~M. Haque},
\newblock \bibinfo{title}{Estimating and comparing response times in
  traditional and connected environments},
\newblock \bibinfo{journal}{Transportation Research Record}
  \bibinfo{volume}{2673} (\bibinfo{year}{2019}) \bibinfo{pages}{674--684}.
\bibitem[{Calvert and van Arem(2020)}]{calvert2020generic}
\bibinfo{author}{S.~Calvert}, \bibinfo{author}{B.~van Arem},
\newblock \bibinfo{title}{A generic multi-level framework for microscopic
  traffic simulation with automated vehicles in mixed traffic},
\newblock \bibinfo{journal}{Transportation Research Part C: Emerging
  Technologies} \bibinfo{volume}{110} (\bibinfo{year}{2020})
  \bibinfo{pages}{291--311}.
\bibitem[{{van Lint} and Calvert(2018)}]{venLint2018generic}
\bibinfo{author}{J.~{van Lint}}, \bibinfo{author}{S.~Calvert},
\newblock \bibinfo{title}{A generic multi-level framework for microscopic
  traffic simulation—theory and an example case in modelling driver
  distraction},
\newblock \bibinfo{journal}{Transportation Research Part B: Methodological}
  \bibinfo{volume}{117} (\bibinfo{year}{2018}) \bibinfo{pages}{63 -- 86}.
  \URLprefix
  \url{http://www.sciencedirect.com/science/article/pii/S0191261518302704}.
  \DOIprefix\doi{https://doi.org/10.1016/j.trb.2018.08.009}.
\bibitem[{Saifuzzaman et~al.(2017)Saifuzzaman, Zheng, Haque, and
  Washington}]{Saifuzzaman2017understanding}
\bibinfo{author}{M.~Saifuzzaman}, \bibinfo{author}{Z.~Zheng},
  \bibinfo{author}{M.~M. Haque}, \bibinfo{author}{S.~Washington},
\newblock \bibinfo{title}{Understanding the mechanism of traffic hysteresis and
  traffic oscillations through the change in task difficulty level},
\newblock \bibinfo{journal}{Transportation Research Part B: Methodological}
  \bibinfo{volume}{105} (\bibinfo{year}{2017}) \bibinfo{pages}{523 -- 538}.
  \URLprefix
  \url{http://www.sciencedirect.com/science/article/pii/S0191261517300474}.
  \DOIprefix\doi{https://doi.org/10.1016/j.trb.2017.09.023}.
\bibitem[{Hamdar et~al.(2015)Hamdar, Mahmassani, and
  Treiber}]{hamdar2015behavioral}
\bibinfo{author}{S.~H. Hamdar}, \bibinfo{author}{H.~S. Mahmassani},
  \bibinfo{author}{M.~Treiber},
\newblock \bibinfo{title}{From behavioral psychology to acceleration modeling:
  Calibration, validation, and exploration of drivers’ cognitive and safety
  parameters in a risk-taking environment},
\newblock \bibinfo{journal}{Transportation Research Part B: Methodological}
  \bibinfo{volume}{78} (\bibinfo{year}{2015}) \bibinfo{pages}{32--53}.
\bibitem[{Sharma et~al.(2019)Sharma, Zheng, Bhaskar, and
  Haque}]{sharma2019modelling}
\bibinfo{author}{A.~Sharma}, \bibinfo{author}{Z.~Zheng},
  \bibinfo{author}{A.~Bhaskar}, \bibinfo{author}{M.~M. Haque},
\newblock \bibinfo{title}{Modelling car-following behaviour of connected
  vehicles with a focus on driver compliance},
\newblock \bibinfo{journal}{Transportation research part B: methodological}
  \bibinfo{volume}{126} (\bibinfo{year}{2019}) \bibinfo{pages}{256--279}.
\bibitem[{Fuller(2011)}]{fuller2011driver}
\bibinfo{author}{R.~Fuller},
\newblock \bibinfo{title}{Driver control theory: From task difficulty
  homeostasis to risk allostasis},
\newblock in: \bibinfo{booktitle}{Handbook of traffic psychology},
  \bibinfo{publisher}{Elsevier}, \bibinfo{year}{2011}, pp.
  \bibinfo{pages}{13--26}.
\bibitem[{Saifuzzaman et~al.(2015)Saifuzzaman, Zheng, Haque, and
  Washington}]{saifuzzaman2015revisiting}
\bibinfo{author}{M.~Saifuzzaman}, \bibinfo{author}{Z.~Zheng},
  \bibinfo{author}{M.~M. Haque}, \bibinfo{author}{S.~Washington},
\newblock \bibinfo{title}{Revisiting the task--capability interface model for
  incorporating human factors into car-following models},
\newblock \bibinfo{journal}{Transportation research part B: methodological}
  \bibinfo{volume}{82} (\bibinfo{year}{2015}) \bibinfo{pages}{1--19}.
\bibitem[{Bibeka et~al.(2019)Bibeka, Songchitruksa, and
  Zhang}]{Bieka2019assessing}
\bibinfo{author}{A.~Bibeka}, \bibinfo{author}{P.~Songchitruksa},
  \bibinfo{author}{Y.~Zhang},
\newblock \bibinfo{title}{Assessing environmental impacts of ad-hoc truck
  platooning on multilane freeways},
\newblock \bibinfo{journal}{Journal of Intelligent Transportation Systems}
  \bibinfo{volume}{0} (\bibinfo{year}{2019}) \bibinfo{pages}{1--12}.
  \DOIprefix\doi{10.1080/15472450.2019.1608441}.
\bibitem[{Killat et~al.(2007)Killat, Schmidt-Eisenlohr, Hartenstein,
  R{\"o}ssel, Vortisch, Assenmacher, and Busch}]{killat2007enabling}
\bibinfo{author}{M.~Killat}, \bibinfo{author}{F.~Schmidt-Eisenlohr},
  \bibinfo{author}{H.~Hartenstein}, \bibinfo{author}{C.~R{\"o}ssel},
  \bibinfo{author}{P.~Vortisch}, \bibinfo{author}{S.~Assenmacher},
  \bibinfo{author}{F.~Busch},
\newblock \bibinfo{title}{Enabling efficient and accurate large-scale
  simulations of vanets for vehicular traffic management},
\newblock in: \bibinfo{booktitle}{Proceedings of the fourth ACM international
  workshop on Vehicular ad hoc networks}, \bibinfo{organization}{ACM},
  \bibinfo{year}{2007}, pp. \bibinfo{pages}{29--38}.
\bibitem[{{Jiang} et~al.(2007){Jiang}, {Chen}, and
  {Delgrossi}}]{Jiang2007communication}
\bibinfo{author}{D.~{Jiang}}, \bibinfo{author}{Q.~{Chen}},
  \bibinfo{author}{L.~{Delgrossi}},
\newblock \bibinfo{title}{Communication density: A channel load metric for
  vehicular communications research},
\newblock in: \bibinfo{booktitle}{2007 IEEE International Conference on Mobile
  Adhoc and Sensor Systems}, \bibinfo{year}{2007}, pp. \bibinfo{pages}{1--8}.
  \DOIprefix\doi{10.1109/MOBHOC.2007.4428734}.
\bibitem[{Killat and Hartenstein(2009)}]{Killat2009}
\bibinfo{author}{M.~Killat}, \bibinfo{author}{H.~Hartenstein},
\newblock \bibinfo{title}{An empirical model for probability of packet
  reception in vehicular ad hoc networks},
\newblock \bibinfo{journal}{EURASIP Journal on Wireless Communications and
  Networking} \bibinfo{volume}{2009} (\bibinfo{year}{2009})
  \bibinfo{pages}{721301}. \URLprefix
  \url{https://doi.org/10.1155/2009/721301}.
  \DOIprefix\doi{10.1155/2009/721301}.
\bibitem[{Leidos(2016)}]{STOLT4}
\bibinfo{author}{Leidos}, \bibinfo{title}{Simulation of Evolutionary
  Introduction of Cooperative Adaptive Cruise Control Equipped Vehicles into
  Traffic}, \bibinfo{type}{Technical Report}, Saxton Transportation Operations
  Laboratory, \bibinfo{year}{2016}.
\bibitem[{Li et~al.(2019)Li, Ma, and Hale}]{Li2019High}
\bibinfo{author}{T.~Li}, \bibinfo{author}{J.~Ma}, \bibinfo{author}{D.~K. Hale},
\newblock \bibinfo{title}{High-occupancy vehicle lanes on the right: an
  alternative design for congestion reduction at freeway merge, diverge, and
  weaving areas},
\newblock \bibinfo{journal}{Transportation Letters} \bibinfo{volume}{0}
  (\bibinfo{year}{2019}) \bibinfo{pages}{1--13}.
  \DOIprefix\doi{10.1080/19427867.2019.1584347}.
\bibitem[{Xiao et~al.(2019)Xiao, Wang, and van Arem}]{xiao2019traffic}
\bibinfo{author}{L.~Xiao}, \bibinfo{author}{M.~Wang}, \bibinfo{author}{B.~van
  Arem},
\newblock \bibinfo{title}{Traffic flow impacts of converting an hov lane into a
  dedicated cacc lane on a freeway corridor},
\newblock \bibinfo{journal}{IEEE Intelligent Transportation Systems Magazine}
  \bibinfo{volume}{12} (\bibinfo{year}{2019}) \bibinfo{pages}{60--73}.
\bibitem[{{Transportation Research Board}(2018)}]{NCHRP}
\bibinfo{author}{{Transportation Research Board}}, \bibinfo{title}{Dedicating
  Lanes for Priority or Exclusive Use by Connected and Automated Vehicles},
  \bibinfo{publisher}{The National Academies Press},
  \bibinfo{address}{Washington, DC}, \bibinfo{year}{2018}. \URLprefix
  \url{https://www.nap.edu/catalog/25366}. \DOIprefix\doi{10.17226/25366}.
\bibitem[{Goodman(1954)}]{goodman1954kolmogorov}
\bibinfo{author}{L.~A. Goodman},
\newblock \bibinfo{title}{Kolmogorov-smirnov tests for psychological
  research.},
\newblock \bibinfo{journal}{Psychological bulletin} \bibinfo{volume}{51}
  (\bibinfo{year}{1954}) \bibinfo{pages}{160}.
\bibitem[{Young(1977)}]{young1977proof}
\bibinfo{author}{I.~T. Young},
\newblock \bibinfo{title}{Proof without prejudice: use of the
  kolmogorov-smirnov test for the analysis of histograms from flow systems and
  other sources.},
\newblock \bibinfo{journal}{Journal of Histochemistry \& Cytochemistry}
  \bibinfo{volume}{25} (\bibinfo{year}{1977}) \bibinfo{pages}{935--941}.
\bibitem[{Li et~al.(2020)Li, Ma, and Hale}]{li2020high}
\bibinfo{author}{T.~Li}, \bibinfo{author}{J.~Ma}, \bibinfo{author}{D.~K. Hale},
\newblock \bibinfo{title}{High-occupancy vehicle lanes on the right: an
  alternative design for congestion reduction at freeway merge, diverge, and
  weaving areas},
\newblock \bibinfo{journal}{Transportation letters} \bibinfo{volume}{12}
  (\bibinfo{year}{2020}) \bibinfo{pages}{233--245}.

\end{thebibliography}
\end{document}